\documentclass[sigconf, nonacm]{acmart}
\usepackage[ruled,linesnumbered,noend]{algorithm2e}
\usepackage{color}
\usepackage{mathtools}
\usepackage{subfigure}
\usepackage[flushleft]{threeparttable}
\usepackage{tablefootnote}

\SetKwFor{ParaFor}{parallel for}{do}{end}

\begin{document}

\title{Graph-based Approximate NN Search: A Revisit}

\author{Hui Wang}
\authornote{Contribution during his intership at NVIDIA}
\affiliation{%
  \institution{Xiamen University}
  \city{Xiamen}
  \country{China}
}
\email{hwang2019@stu.xmu.edu.cn}

\author{Yong Wang}
\affiliation{%
  \institution{NVIDIA}
  \city{Shanghai}
  \country{China}
}
\email{yongw@nvidia.com}

\author{Wan-Lei Zhao}
\authornote{Corresponding author}
\affiliation{%
  \institution{Xiamen University}
  \city{Xiamen}
  \country{China}
}
\email{wlzhao@xmu.edu.cn}

\begin{abstract}
Nearest neighbor search plays a fundamental role in many disciplines such as multimedia information retrieval, data-mining, and machine learning. The graph-based search approaches show superior performance over other types of approaches in recent studies. In this paper, the graph-based NN search is revisited. We optimize two key components in the approach, namely the search procedure and the graph that supports the search. For the graph construction, a two-stage graph diversification scheme is proposed, which makes a good trade-off between the efficiency and reachability for the search procedure that builds upon it. Moreover, the proposed diversification scheme allows the search procedure to decide dynamically how many nodes should be visited in one node's neighborhood. By this way, the computing power of the devices is fully utilized when the search is carried out under different circumstances. Furthermore, two NN search procedures are designed respectively for small and large batch queries on the GPU. The optimized NN search, when being supported by the two-stage diversified graph, outperforms all the state-of-the-art approaches on both the CPU and the GPU across all the considered large-scale datasets.
\end{abstract}

  


\maketitle

\section{Introduction}
Nearest neighbor search is a fundamental issue that arises from several disciplines such as database, information retrieval, pattern recognition, and machine learning. Given a query ($q \in S^d$) and a distance metric $m(\cdot,\cdot)$, the search procedure returns its nearest neighbor or its top-\textit{k} nearest neighbors (\textit{k}-NN) from a candidate set $C=\{x|x \in S^d\}$, where \textit{d} is the data dimension. The early study about the \textit{k}-NN search issue could be traced back to the \textit{1970s} since the proposal of B-tree~\cite{btree:commer79}. After the continuous exploration in a half century, efficient solutions are available for some of the sub-issues. For instance, KD-tree~\cite{kdtree75} works well for low-dimensional data. In the high and sparse dimensional case, the inverted file is one of the most efficient indexing structure for data such as textual documents and web pages, which is still the core structure in nowadays search engines. Nevertheless, for high-dimensional (\textit{e.g.}, $d > 20$) and dense data, this issue remains challenging due to the widely known ``the curse of dimensionality''. In recent years, this issue becomes more and more imminent given that the big data emerge in various forms across different fields.

In the last three decades, the research on NN search has been pushed forward by different waves of data indexing demands. There are in general four categories of approaches that have been proposed. Namely, they are tree-based approaches such as R-tree~\cite{rtree84}, hash approaches~\cite{datar2004locality,mlsh07}, quantization-based approaches~\cite{jegou2010product,pami14:flann}, and graph-based approaches~\cite{malkov2020efficient, fu2019fast}. Amongst all these approaches, graph-based approaches show the highest performance according to recent studies~\cite{li2019approximate}. In general, it is an $A^*$-like search procedure. The search traverses iteratively over a pre-built approximate \textit{k}-NN graph or a relative neighborhood graph (RNG)~\cite{ChenW18} by the best-first search. The search proceeds to the next stage by expanding the neighbors of the closest sample in the rank-list. It ascends closer to the true nearest neighbor in each round until no better candidates could be found.

In the graph-based approaches, the search procedure is undertaken following a path from a random seed to the neighborhood of the target sample. The path is on a manifold that is formed by the graph. In many real scenarios, the dimension of this manifold is much lower than the data dimension. The search actually explores a much lower dimensional space. In the recent studies~\cite{li2019approximate, malkov2020efficient}, the search efficiency is further boosted by diversifying the \textit{k}-NN graph into an approximation of the relative neighborhood graph. Although different diversifying schemes are proposed in~\cite{li2019approximate, malkov2020efficient,fu2019fast}, they are largely out of the same principle. Namely, the query tries to avoid the comparison with the samples whose close neighbors have been already compared. For most of the diversification schemes, the edges directing to these redundant neighbors are simply removed. However, this is not always appropriate. On the one hand, less number of comparisons leads to the higher efficiency of the search. On the other hand, more number of visits to different samples in the graph increases the likelihood of reaching a true nearest neighbor. 

In this paper, we optimize not only the diversification scheme that is applied on the \textit{k}-NN graph but the NN search procedures that build upon the graph. The contributions are at least two folds.
\begin{itemize}
	\item {A two-stage graph diversification scheme is proposed, which makes a good balance between the connectivity and the sparsity of the graph. It outperforms all the other graphs when it is used to support different variants of the search procedure on both the GPU and the CPU.}
	\item {Moreover, two different NN search procedures are designed on the GPU for different scales of batch queries. The parallelism power of the NN search on the GPU is optimized for both small and large batch queries. As will be revealed in the experiment, our approaches outperform the state-of-the-art GPU-based approaches considerably. In particular, for queries that arrive in small batches, our NN search is shown to be \textit{2$\sim$10} times faster than the state-of-the-art approaches.}
\end{itemize}


\section{Related Work}
\subsection{Non-Graph based Approaches}
There are generally four major categories of NN search approaches in the literature. In the \textit{1980}s and early \textit{1990}s, most of the approaches are an extension of B-tree. The representative indexing trees  are KD-tree~\cite{kdtree75}, R-tree~\cite{rtree84}, and X-tree~\cite{vldb96:xtree}, etc. Essentially, the idea behind these approaches is similar. They partition the space into hyper-rectangles under different strategies. Close neighbors are assumed to reside under the same hierarchy of hyper-rectangles. However, the situation is much more complicated in the high-dimensional space. The partition scheme does not exclude the possibility that the nearest neighbors reside outside of one branch. It becomes inevitable to probe extensively over a large number of neighboring branches. Quantization-based approaches~\cite{pami14:flann} partition the space in an alternative way. The candidate data are quantized with respect to a pre-defined code-book~\cite{iccv03:sivic} or a hierarchy of code-books~\cite{pami14:flann}. Different from tree-based approaches, the space is partitioned by a series of \textit{Voronoi} cells, each of which is regularized by one vocabulary word in the code-book. These two types of approaches share one thing in common. Namely, the query is compared to the indexing nodes (non-leaf nodes in the tree or words in the code-book) to locate the nearest neighbor. 

In recent works, several attempts have been made to compress the candidate samples by vector quantization~\cite{babenko2016efficient,icml14:tzhang, sq14, tit06:gray06} or sub-vector quantization~\cite{jegou2010product,ge2013optimized}. On the one hand, the words in the code-books are used to guide the NN search. On the other hand, several words in combination are used to approximate the candidate samples. The NN search is conducted between the query and the compressed candidate samples. The distances  between the query and the candidate samples are approximated by the distances between the query and the code words. One advantage of these approaches is that it is no need to keep raw vectors in the memory. However, due to the compression on the candidate samples, high search quality is hardly achievable. Furthermore, these types of approaches are usually only feasible for metric spaces of $\textit{l}_p$-norm.

Apart from the above two types of approaches, hash approaches have been introduced to the nearest neighbor search in~\cite{datar2004locality,mlsh07}. The samples are mapped into hash codes by locality sensitive hashing (LSH). Close neighbors in the vector space are expected to share the same or similar hash codes. Therefore, the query will only be compared to the samples with the same or similar hash codes to find out the nearest neighbors. The high compression rate leads to low memory cost and efficient comparison, however, low search quality as well. Moreover, compared to the compression by vector quantization, comparisons with the raw vectors are still necessary, which in turn requires the maintaining of all the raw vectors in the memory. Moreover, it is non-trivial to define a generic hash mapping scheme that works for various types of data.

\subsection{Graph-based Approaches}
For all the aforementioned approaches, the candidate samples have been indexed by non-leaf nodes of a tree, vocabulary words or hash codes. The search localizes the close neighbors via a series of comparisons between the query and these indexing nodes. Different from the aforementioned approaches, every sample in the graph-based approaches acts as an indexing node. In general, two stages are involved in the graph-based search. In the first step, a \textit{k}-NN graph or a relative neighborhood graph is constructed. Each sample (node) links to its close neighbors. This is usually fulfilled offline. In the second step, the search starts from comparing with a group of random samples on the graph. By expanding the links to the neighbors of visited nodes (being compared), the search ascends closer to the true neighbors in each round. The search iterates until no better candidates could be found. Approaches in~\cite{icai11:kiana,malkov2020efficient,fu2019fast,arxiv22:fabian} follow a similar search procedure. The major difference between them lies in the graph used to support the search. According to the recent studies~\cite{li2019approximate,prokhorenkova2020graph}, NN search that is supported by the \textit{k}-NN graph already outperforms most of other types of approaches. Another advantage of graph-based approaches over other types of approaches is that they are feasible for different distance metrics.

According to recent studies~\cite{malkov2020efficient,fu2019fast,li2019approximate}, the search performance can be further boosted when the supporting \textit{k}-NN graph is further diversified. Although different variants of diversification approaches have been proposed and their performance ranking varies across different datasets, they are shown to be essentially similar~\cite{wang2021comprehensive}. The major principle out of them is to remove the edges that are shorter to other neighbors in one node's neighborhood than they are to the node. These edges are mostly directed to the nodes already reachable via the reserved edges. The diversification helps to reduce the redundant comparisons during the NN search. Most of the graph diversification schemes such as DPG~\cite{li2019approximate}, SSG~\cite{fu2021high}, and GD in HNSW~\cite{malkov2020efficient} are applied on an approximate \textit{k}-NN graph. As an exception, NSG~\cite{fu2019fast} applies diversification on a searching route from a random seed to a node. The performance of these schemes varies considerably across different datasets, where the data distribution changes. 

In this paper, we are going to propose a two-stage graph diversification scheme. The first stage diversification helps to remove those apparently redundant edges. The second stage diversification associates an occlusion factor with each edge, which indicates the importance of an edge. For different search procedures, they could choose to either visit or skip an edge according to this factor under different NN search circumstances. When computing resources are sufficient for one search, edges even with high occlusion factors can be visited. Such that we aim to maximize the search efficiency that we can reach on one device.

\subsection{NN Search on the GPU}
On the CPU, the most computationally intensive operation in the NN search is the distance computation. For all the above approaches designed for the CPU, the motivation is to find out the nearest neighbors while limiting the number of distance computations as few as possible. On the GPU, the situation is different, where high parallel computing performance is available. The distance computation is relatively cheap in this circumstance. For this reason, the high efficiency is achieved even by performing exhaustive comparisons between the query and the whole candidate set according to~\cite{johnson2019billion}.  The major issue lies in the throughput mismatch between the GPU cores and the different levels of memories. In~\cite{johnson2019billion}, the nearest neighbors are found by merging the results from exhaustive comparison by \textit{k}-selection on each thread block. However, it is wasteful to perform the exhaustive comparisons for NN search~\cite{sigir21:gaips}. In recent studies~\cite{sigir21:gaips, arxiv22:fabian, zhao2020song}, the NN search is carried out with the support of graph on the GPU. The query is expected to compare only with samples along a search path. In order to make the full use of GPU computing power, the distances between the query and the samples in one NN list are computed on multiple threads of thread blocks~\cite{zhao2020song}. Additionally, the bloom filter is adopted to check whether a sample has been visited before~\cite{zhao2020song} to avoid repetitive computation.

Unfortunately, approaches in~\cite{arxiv22:fabian, zhao2020song} are designed for queries that arrive in large batches. When the queries arrive in small batches, the advantage of GPU's computing power cannot be fully utilized. In this paper, two NN search algorithms are designed for the queries that arrive in different scales of batches on the GPU. When these search algorithms are supported by the proposed graph, they both perform better than the state-of-the-art approaches.

\section{Two-Stage Graph Diversification}
\label{sec:fdg}
\subsection{Preliminaries for Graph Diversification}
Most of the graph-based approaches follow a similar search procedure, the major difference among them lies in the graph used to support the search. The graph is usually diversified from an approximate \textit{k}-NN graph. The diversification aims to keep only a small portion of the edges in \textit{k}-NN graph. Given there are \textit{k} edges in one \textit{k}-NN list, several of them are directed to the same or similar group of nodes in the graph. On the one hand, it is unnecessary to keep all of them since the comparison with all of the directed samples makes no improvement for the NN search. The removal of such edges reduces the redundant comparisons. On the other hand, certain amount edges should be kept in order to maintain the connectivity of the graph. Since all these strategies in the literatures (FANNG~\cite{harwood2016fanng}, HNSW~\cite{malkov2020efficient}, GD~\cite{tbd21:zhao}, DPG~\cite{li2019approximate} and NSG~\cite{fu2019fast}) are similar, we will review the strategy used in HNSW only in the following.
  
For each node in the graph, the diversification is performed on a group of edges that are directed to its neighbors. In most of the cases~\cite{malkov2020efficient,tbd21:zhao,li2019approximate}, the edges directed to its \textit{k}-NNs are treated as the candidates. Alternatively, the edges on the searching route from a random sample to the node are treated as the candidates~\cite{fu2019fast}. The edges are kept when they are not occluded. Given current node $x_0$, and its two neighbors $x_i$ and $x_j$, edge$\langle x_0,x_j \rangle$ is occluded by edge$\langle x_0,x_i \rangle$ if the following two inequalities hold\footnote{Without loss of generality, we assume the smaller the distance between two samples, the closer they are.}

\begin{equation}
\left\lbrace \begin{array}{l}
m(x_0, x_i) < m(x_0, x_j) \\
m(x_i, x_j) < m(x_0, x_j)
\end{array}. \right.
\label{eqn:gd}
\end{equation}
In this case, the occluded edge$\langle x_0, x_j \rangle$ will be discarded. The occlusion relation between neighbors is checked starting from the closest neighbor. According to the above rule, the closest neighbor is the first to be selected into the diversified list. While the remaining edges are checked pairwise with all the edges that are already put into the diversified list.

\begin{figure}[t]
    \centering
    \subfigure[]{
       \includegraphics[width=0.43\linewidth]{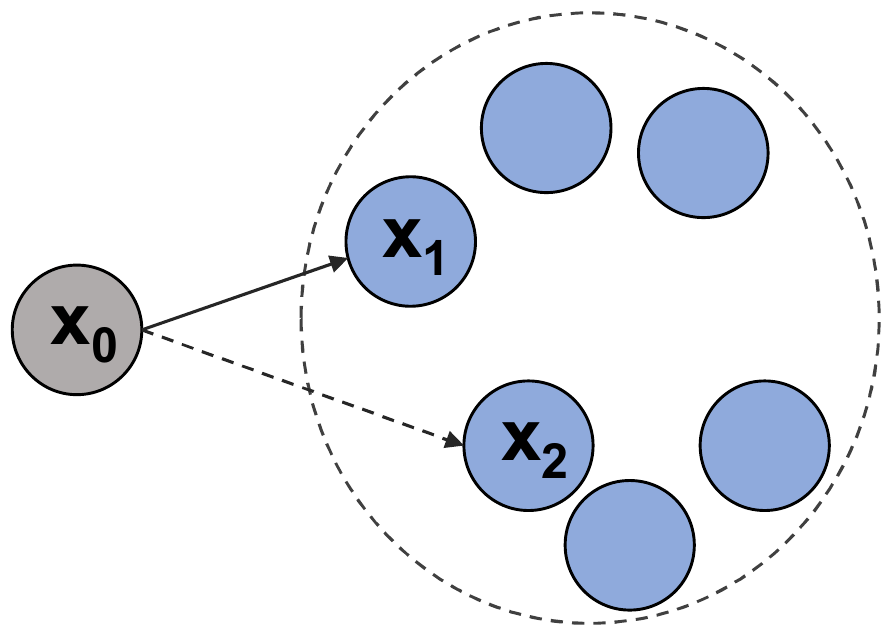}
       \label{fig:gd}
    }
    \subfigure[]{
       \includegraphics[width=0.43\linewidth]{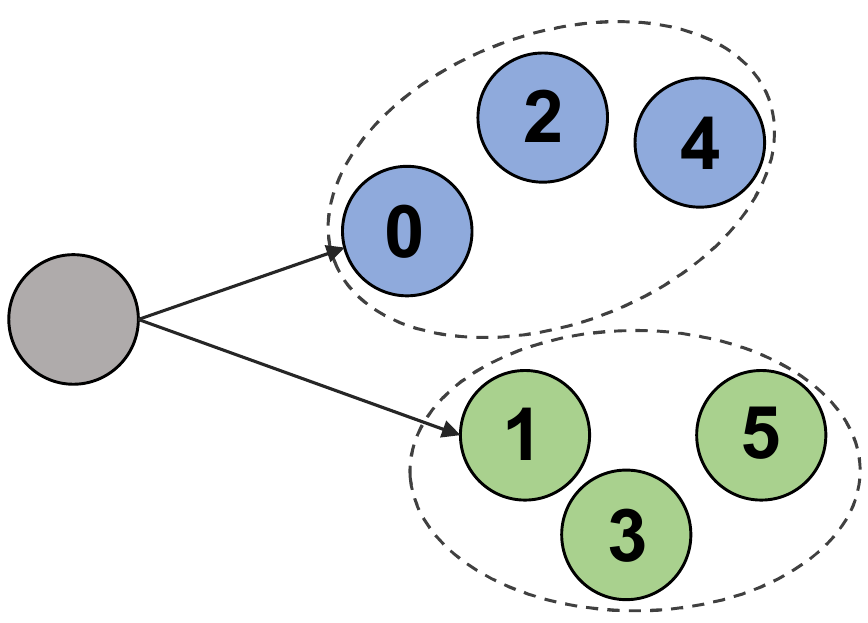}
       \label{fig:fdg}
    }
    \caption{An illustration of the occlusion phenomenon. The dashed line indicates a removed edge and the numbers in (b) indicate the occlusion factors. Two edges  in (b) are retained because their occlusion factors are below \textit{2}. }
    \label{fig:occlusion}
\end{figure}

As shown in \autoref{fig:occlusion}\ref{sub@fig:gd}, edge$\langle x_0, x_2 \rangle$ is discarded because it is occluded by edge$ \langle x_0, x_1 \rangle$. Moreover, the edges from $x_0$ to other blue nodes are all occluded by edge$\langle x_0, x_1 \rangle$. These blue nodes actually form a cluster. According to~\cite{malkov2020efficient}, it is unnecessary to visit edges other than edge$\langle x_0, x_1 \rangle$ as they offer similar guiding information. Compared to the edges that have been discarded, the kept edges are more important to the search. Similar to the description in DPG~\cite{li2019approximate}, this strategy diversifies the direction of the edges pointing out from each node, so we call it graph diversification (GD)~\cite{tbd21:zhao}.

In $l_2$-space, the graph diversification strategy ensures that the angle between edges is at least $60^{\circ}$. One potential issue about GD is that the edges connected to far away clusters are simply discarded when there could be multiple clusters within the range of $60^{\circ}$. As illustrated in~\autoref{fig:occlusion}\ref{sub@fig:fdg}, the green nodes and the blue nodes come from different clusters, only one edge will be kept for each cluster under the GD rule. For this reason, the reachability of the nodes inside the cluster is degraded.

We can learn from the above example that the removal of an edge is a tricky trade-off. We should consider both the cost of visiting an edge and the guiding information it supplies. On the one hand, keeping the edges connected to smaller scale clusters improves the connectivity between nodes, which in turn improves the reachability of the search procedure to these clusters. On the other hand, it induces more distance computations per hop as well. 

In addition, for different types of search, the costs spent on distance computations are different. For small batch search on the GPU, the distance computation can be carried out in high parallel. More edge connections in the graph are affordable. In contrast, for NN search in large batches on the GPU, the search has to be processed in a quite different manner. More number of edge connections increases the computation cost considerably. However, one cannot maintain several graphs with different levels of sparsity. Maintaining several graphs in the memory to support different search batches leads to several times larger memory consumption. Given the size of dataset is big, this is simply not affordable, in particular for the GPU. As a consequence, it is expected to only maintain one diversified graph to support NN search in different batch sizes. In order to maintain the full speed for different search tasks, the diversified graph should allow the search process to choose which edges should be visited or skipped under different circumstances.

In the following section, a two-stage diversification strategy is presented. Accordingly, the graph diversified by our approach is referred to as Two-stage Diversified Graph (TSDG). Different from most of the existing approaches, a portion of the edges within some clusters are kept even they are occluded. An occlusion factor is associated with each edge. Search procedure is allowed to choose either to visit or to skip these edges by the factor under different scenarios.

\subsection{Relaxed Graph Diversification}
On the one hand, graph diversification reduces the average number of comparisons per hop. On the other hand, it decreases the connectivity in the graph, which increases the necessary number of hops before the query reaches its nearest neighbor. In order to make a balance between these two competing properties of a graph, we propose to undertake the diversification in two stages.

In the first stage, similar to~\cite{subramanya2019diskann}, a relaxed graph diversification is performed on each NN list of \textit{k}-NN graph. Namely, the occlusion condition is defined as
\begin{equation}
\left\lbrace \begin{array}{l}
\alpha{\cdot}m(x_0, x_i) < m(x_0, x_j) \\
\alpha{\cdot}m(x_i, x_j) < m(x_0, x_j)
\end{array}, \right.
\label{eqn:sgd}
\end{equation} 
where $\alpha$ is a parameter usually greater than \textit{1.1}. According to our observation, only \textit{6$\sim$26\%} of the edges are left after this operation\footnote{The statistics are conducted on six large-scale datasets that are adopted in the experiments.}. Compared to the graph diversified by the original GD, more edges are kept. These extra edges are expected to maintain the connections within local clusters. However, the data distributions vary from one dataset to another. Such a less sparsified graph does not provide the flexibility of visiting certain edges according to their importance under different circumstances. In the next stage, we will address this issue by soft graph diversification. 

\subsection{Soft Graph Diversification}
The result of the first stage GD is a sparsified \textit{k}-NN graph. Before we proceed with the \textit{2nd}-stage diversification, the reverse edges of each node are appended to the sparsified \textit{k}-NN list. After this operation, the graph is transformed into an undirected graph. Thereafter, the second round of diversification is undertaken on the NN list of each node. For each edge that survives the first round of diversification, it is associated with an occlusion factor $\lambda$, which is initialized to \textit{0}. We check whether it is occluded by any other edge with Inequation~\ref{eqn:gd}. For instance, if the inequalities hold, the occlusion factor of edge$\langle x_0, x_j \rangle$ is incremented by \textit{1}. After all the occlusion factors are calculated, edges are sorted by the occlusion factors  in ascending order. If two edges share the same occlusion factor, they are further sorted by their distances (to node $x_0$) in ascending order. Edges whose occlusion factors are greater than a threshold $\lambda_0$ are simply removed. Taking \autoref{fig:occlusion}\ref{sub@fig:fdg} as an example, we keep the edges with occlusion factor \textit{1} and below, therefore an extra edge will point to samples in the local cluster (green nodes). This is the major difference from the original GD rule. 

The occlusion factor provides a means of measuring the importance of edges. Namely, the lower is this value, the more important of this edge is. This is important when NN search runs on different devices or for different search tasks. For the search on the GPU that we will describe later, small batch queries perform better on graphs with high average degree. While for large batch queries,  one query is supposedly fulfilled by one GPU warp. The cost of distance computation is much higher than the cost of data structure maintenance in the memories in one warp. It is therefore inefficient to perform search on a graph with extra large degree, which induces more number of distance computations.  The graph under soft diversification allows the search procedure to choose how many samples in the list to visit. It therefore no need to maintain several graphs in different average degrees for different batch queries' use. In the large batch size case, only the top-ranked edges are considered. 
%

\begin{figure}[t]
    \centering
    \subfigure[]{
       \includegraphics[width=0.3\linewidth]{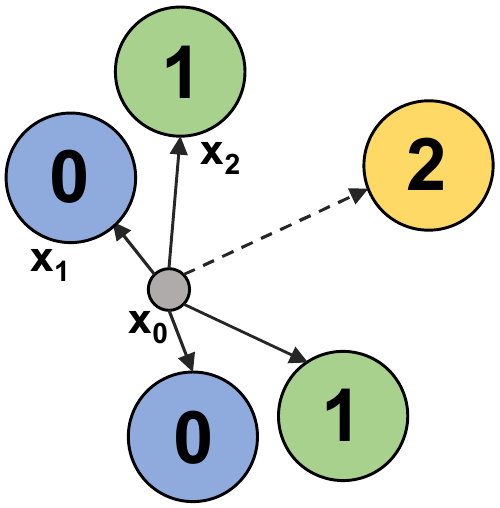}
       \label{fig:dpg_bad}
    }
    \hspace{0.1\linewidth}
    \subfigure[]{
       \includegraphics[width=0.3\linewidth]{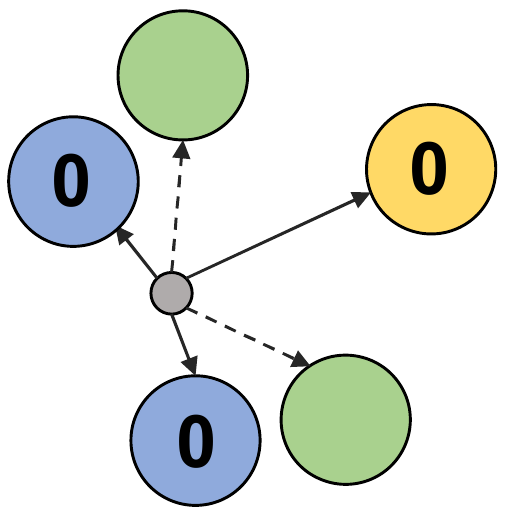}
       \label{fig:dpg_good}
    }
    \caption{The illustration of two-stage diversification. The numbers on the nodes are the occlusion factors associated with the corresponding edges. Figure (a) illustrates the scenario that only the \textit{2nd}-stage diversification is applied. The dashed edge will be removed according to the rule. Figure (b) illustrates the two-stage diversification. The edges connecting to the green nodes are removed by the \textit{1st}-stage diversification. The edge connecting to the yellow node is retained.}
    \label{fig:two_stage}
\end{figure}

To this end, it seems that the first stage diversification is redundant. One can apply the second stage diversification on the \textit{k}-NN graph directly. However, there are two reasons that make the first stage diversification still necessary. First of all, the first stage diversification helps to prune the edges which would be associated with unexpectedly low $\lambda$ according to the soft GD rule. This is demonstrated in Figure~\ref{fig:two_stage}. As shown in the figure, sample $x_2$ is very close to $x_1$ while far from the rest. In this case, edge$\langle x_0,x_2 \rangle$ is only occluded by edge$\langle x_0,x_1 \rangle$. According to our rule, its $\lambda$ is only \textit{1}. However, according to the principle of diversification, it should not be kept as it is very close to $x_1$. The first stage diversification helps to discard such edges that the second stage cannot assign a factor properly. Furthermore, the size of \textit{k}-NN list could be large in some scenarios, say several hundreds. The size of NN list could be even much larger when the reverse edges are appended. Performing soft diversification on such a list could be time-consuming. The first stage diversification helps to prune the edges that the second stage should not consider. Therefore, the first stage diversification reduces the overall time cost of the second stage graph diversification.

The two-stage diversification can be carried out efficiently on the GPU. The calculation of the occlusion factors for the edges of different nodes is completely independent and can be easily parallelized on the GPU. Finally, unlike NSG~\cite{fu2019fast} or Vamana~\cite{subramanya2019diskann} which obtain candidate edges from the searched routes, the candidate edges are the edges from the \textit{k}-NN graph, so we can load nodes and candidate edges into the GPU in batches, which reduces the GPU memory usage significantly.

\begin{figure}[t]
    \centering
    \includegraphics[width=0.8\linewidth]{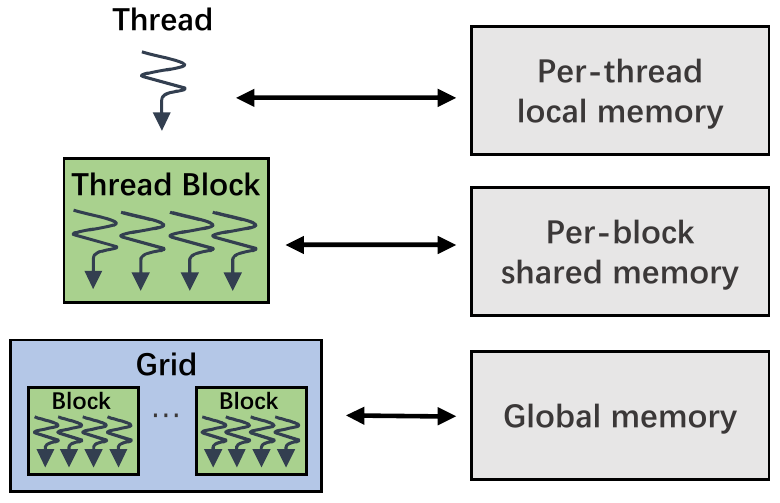}
    \caption{Simplified CUDA programming model.}
    \label{fig:cuda_model}
\end{figure}

\section{Graph-based ANN Search on the GPU}
Once the diversified graph is built, one can proceed with the graph-based NN search. Most of the existing approaches on the CPU are the variants of hill-climbing procedure~\cite{icai11:kiana}. Their performance is similar. In our implementation, we adopt the procedure from NSG~\cite{fu2019fast} for NN search on the CPU. While the existing search procedures on the GPU~\cite{zhao2020song,arxiv22:fabian} are designed mostly for the large batch queries, where thousands of queries are processed in parallel. Such kind of design leads to poor performance when the queries arrive in small batches. In this section, the optimized NN search procedures on the GPU are presented for small and large batch queries respectively. Before we present our search procedures, a brief review about the programming model of CUDA is given.

The GPU is designed to excel at executing thousands of threads in parallel. In the CUDA programming model, threads are logically divided into \textit{1}, \textit{2}, or \textit{3} dimensional groups referred to as thread \textit{blocks}. These thread blocks are further grouped into a \textit{1}, \textit{2} or \textit{3} dimensional \textit{grid}. A function that is to be run on the GPU is called as a \textit{kernel}. A grid of thread blocks are launched for one kernel. In CUDA, threads are created, managed and executed in groups of \textit{32} parallel threads called \textit{warps}. So each thread block can be partitioned into multiple warps. In the graph-based NN search, almost all the data structures we use are small in scale and therefore are maintained in \textit{shared memory}, which is a low-latency memory but only has limited capacity. Each thread block has shared memory visible to all threads in it. 

Given a batch of queries arrive, a straightforward way to undertake the search on the GPU is to run each individual query on one block. As there are thousands of cores run in parallel, this process could be very efficient. However, it will not be as efficient as we expected when the batch size is small. In this case, the number of thread blocks is much greater than the number of queries. There would be many SMs remain idle. To address this issue, we fulfill the parallel NN search on the GPU differently when the queries arrive in small and large batches. 

The division between the small and large batches is closely related to the available GPU resources and data dimension. In general, the more streaming multiprocessors (SM) are in a GPU device, the query batch should be larger if we want to fully utilize its computing power. The empirical formula for the division threshold between the small and large batches is $\frac{a \cdot SMs + b}{d}$, where $a$ and $b$ are device-related constants. $SMs$ is the number of SMs, and $d$ is the dimensionality of the data. Take our case NVIDIA GeForce RTX 3090 as an example, which has \textit{82} SMs and \textit{128} CUDA cores per SM. On SIFT1M dataset (\textit{128} dimensions), the threshold between small and large batches is about \textit{300}, while the threshold on GIST1M dataset (960 dimensions) is about \textit{150}.


\subsection{Small Batch Search on the GPU}
When the queries arrive in small batches, many GPU SMs remain idle if one query is assigned to one thread block inside the GPU. For this reason, the computation power of GPU cannot be fully utilized. To address this issue, the parallelization for small batch queries is realized in a novel way. Given a batch of queries arrive  $Q=\{q_1,{\cdots}, q_i,\cdots, q_m\}$, one query $q_i$ is fulfilled on multiple thread blocks on the GPU. For each $q_i$, a group of simple greedy searches $S=\{s_1,{\cdots}, s_j,\cdots, s_{t_0}\}$, where ${t_0}$ is the number of searches, are launched on $t_0$ thread blocks. Each search procedure $s_j$ runs independently and returns a ranking $R_{ij}$ for $q_i$. This ranking is finally merged into the ranking $R_i$ for query $q_i$. 

\begin{algorithm}[t]
    \caption{NN Search for Small Batch on the GPU}
    \label{alg:simple_search}
    \KwData{graph $G$, query $q_i$, number of hop limit $T$}
    \KwResult{top-\textit{32} approximate nearest neighbors}
    $R_{ij} \leftarrow 32$ ID-distance pairs$(\infty, \infty)$\;
    $U \leftarrow$ \textbf{Generate} 32 random starting nodes from $G$\;
    $u \leftarrow$ \textbf{Get} closest node to $q_i$ in $U$\;
    $improved \leftarrow$ \textbf{true}; $t \leftarrow 0$\;
    \While{improved and $t < T$} {
        $R_{temp} \leftarrow 32$ ID-distance pairs$(\infty, \infty)$\;
        \For{every 32 neighbors $V$ of $u$ in $G$} {
            \label{simple_search:for_1}
            \ParaFor{$v \in V$} {
                $dist \leftarrow m(v, q_i)$\;
                \If{$dist < R_{temp}[warp\_id].dist$}{
                    \label{simple_search:warp_id}
                    $R_{temp}[warp\_id] \leftarrow (v, dist)$\;
                }
            }
        }
        \textbf{Update} $R_{ij}$ with $R_{temp}$\;
        \label{simple_search:update}
        $u \leftarrow$ \textbf{Get} closest node to $q_i$ in $R_{temp}$\;
        \If{$R_{ij}$ is not updated} {
            $improved \leftarrow$ \textbf{false}\;
        }
    }
    \Return{$R_{ij}$}\;
\end{algorithm} 

The procedure of the simple greedy search is summarized in~\autoref{alg:simple_search}. The search starts from the best sample that is selected from \textit{32} random seeds. According to our offline test, this simple seed selection scheme is as effective as the hierarchical strategy in HNSW~\cite{malkov2020efficient} and the ``navigating node'' strategy in NSG~\cite{fu2019fast}. The starting sample is set as the ``current node'' \textit{u}. As shown in \textit{Lines 7-11}, \textit{32} samples from \textit{u}'s NN list are compared with query $q_i$. The number of samples compared here is in line with the number of warps in one thread block. Each distance is computed by a warp in parallel. The resulting \textit{32} distances are used to replace the results kept in $R_{temp}$ if they are smaller. $R_{temp}$ is an unordered array in the shared memory or registers. The size of $R_{temp}$ is \textit{32} as well, which is in line with the number of warps in each block. $R_{temp}$ is accessed by the warp ID. The computed distance from one warp with ID $warp\_id$ will only compare with one cell $R_{temp}[warp\_id]$. 
For this reason, it is not guaranteed that $R_{temp}$ keeps the closest top-\textit{k} samples that are visited so far. However, such ad hoc update only requires the threads to access $R_{temp}$ once, which is very efficient. The smallest sample in $R_{temp}$ is selected as the expanding point \textit{u} for the next round.


Array $R_{ij}$ is used to keep the final top-\textit{k} results for a single search $s_j$. Similar to $R_{temp}$, it is maintained in the shared memory or registers. The update operation in \textit{Line 12} of \autoref{alg:simple_search} is carried out by one warp. It is an incomplete ascending sort of $R_{temp}$ with bitonic sorter~\cite{batcher1968sorting}, ensuring that the first \textit{16} elements are the smallest. The last \textit{16} elements of $R_{ij}$ are replaced with the top-\textit{16} smallest elements of $R_{temp}$. Finally, a complete sort by bitonic sorter is conducted on $R_{ij}$. Since the update operation on $R_{ij}$ is time-consuming, we should try to reduce the number of search hops. As shown by the experiments in \cite{subramanya2019diskann}, the higher average degree of the graph leads to the fewer hops in  NN search. Graph with two-stage diversification matches well to such scenario. It allows the NN search to traverse over a graph with a high average degree by visiting more edges with larger occlusion factors.

As one could see from~\autoref{alg:simple_search}, no expansion queue is adopted. This is for the sake of efficiency. This greedy search converges in simply \textit{4-5} iterations. For this reason, one cannot expect good performance from a single search. The NN search quality is enhanced by running multiple such cheap searches in parallel for one query. 

When performing multiple independent searches for the same query, redundant distance computations may happen. Therefore, the results collected from different searches might be duplicated. Fortunately, due to the poor precision of the single greedy search, the likelihood of duplication between different searches is low. This motivates us to sacrifice the precision of each independent search to improve the search speed. We increase the search quality by increasing the number of independent searches $t_0$. Because the batch size is small, increasing $t_0$ makes little impact on the search speed, while the recall is improved considerably. 

\subsection{Large Batch Search on the GPU}
When the queries arrive in large batch, the search is undertaken in a straightforward way. One thread block runs as a CPU core, and is responsible for NN search of a single query. Thousands of queries can be fulfilled in parallel on thousands of thread blocks. The constructed graph is held in the global memory to support all the thread blocks in the GPU. In each thread block, we assign \textit{32} threads (one warp) work in parallel serving for one query $q_i$. The distance computation, expansion queue update, and the top ranking list maintenance for each query are all operated in parallel by the warp.

The search algorithm performed by each thread block is shown in \autoref{alg:common_search}. The search process is very similar to the bottom-layer search used in HNSW~\cite{malkov2020efficient} on the CPU. While the major difference lies in the maintenance of three key data structures. Namely, they are the top ranking array $R$, the expansion queue $C$, and the look-up table $V$. The top ranking array $R$ is used to keep the top-\textit{k} search results and its size is fixed to \textit{k}. Since the size of $R$ is fixed for efficiency, similar to~\cite{arxiv22:fabian}, we use a threshold $\Delta$ as the termination condition for the search. The expansion queue $C$ keeps the samples whose NN list will be visited in the future rounds. $V$ is used to check whether a sample has been visited before. All these three structures are kept in the shared memory for the sake of efficiency.

\begin{algorithm}[t]
    \caption{NN Search for Large Batch on the GPU}
    \label{alg:common_search}
    \KwData{graph $G$, query $q$, number of hop limit $T$, probe threshold $\Delta$}
    \KwResult{top-\textit{k} approximate nearest neighbors}
    $U \leftarrow$ \textbf{Generate} 32 random starting nodes from $G$\;
    $u \leftarrow$ \textbf{Get} closest node to $q$ in $U$\;
    $R \leftarrow \{u\}$; \hspace{5pt} // priority queue of found nearest neighbors \\
    $C \leftarrow \{u\}$; \hspace{3.5pt} // expansion queue of candidates \\
    $V \leftarrow \emptyset$; \hspace{10.8pt} // set of visited elements \\
    $t \leftarrow 0$\;
    \While{$\vert C \vert > 0$ and $t < T$} {
        $t \leftarrow t + 1$\;
		\textit{u} $\leftarrow$ pop nearest element to $q$ from $C$\;
		\textit{f} $\leftarrow$ get furthest sample to $q$ from $R$\;
		\If{$m(u, q) > m(f, q) + \Delta$} {
            \label{common_search:term_condi}
			break\;
		}
        $V$.Add($u$)\;
        \label{common_search:hash}
		\For{each neighbor $e$ of $u$ in G} {
			\If {$e \notin V$ and $e \notin C$} {
                \label{common_search:visited}
				\textit{f} $\leftarrow$ get furthest sample to $q$ from $R$\;
				\If {$m(e, q) < m(f, q)$ or $|R| < k$} {
                    $R$.Push($e$)\;
					$C$.Push($e$)\;
                    \If {$|R| > k$} {
                        $R$.Pop(); \quad // remove furthest sample \\
                    }
				}
			}
		}
    }
    \Return{$R$}\;
\end{algorithm} 

Similar to the small batch case, the query starts from the closest sample selected from \textit{32} random seeds. This way is as effective as the hierarchical structure in HNSW~\cite{malkov2020efficient} and ``navigating node'' in NSG~\cite{fu2019fast}. In each round of NN search, one sample \textit{u} from $C$ is popped out. The query is compared to its unvisited neighbors \textit{e}. Once sample \textit{e} is compared to the query, it is pushed to $R$ and $C$ respectively. Sample $u$ is pushed to the look-up table $V$. The iteration continues until $C$ is empty or no improvement can be made. Although the above NN search makes no significant difference from the popular graph-based search procedure on the CPU, the design of structures about $R$, $C$, and $V$ is essentially different from its CPU versions. In the following, more details are given to explain how these structures are designed and maintained.

The expansion queue $C$ is implemented as a long array. This long array is further divided into multiple segments with equal size. The segment size is set to the same as the number of threads in a block. In our case, it is \textit{32}. Each segment is structured as a sorted circle array. It is sorted in ascending order according to the distances of samples' to the query. Given sample \textit{e}, the segment to keep it is located by $e.id \% m$, where \textit{m} is the number of segments and $e.id$ is the sample ID. It is, therefore, inserted to the segment $e.id \% m$. The most distant sample in that segment is popped out when the segment is full. All the top-\textit{1} elements from \textit{m} segments are checked when we want to pop out one sample to expand. The smallest is extracted from them and popped out from the corresponding segment. Compared to the priority queue used in~\cite{arxiv22:fabian}, we divide the queue into dozens of segments. The size of each segment is in line with the number of threads in a block. Therefore, the access to segment cells could be fulfilled in one round, which considerably reduces the number of scans and threads.

The look-up table $V$ is implemented as a long array and divided into \textit{m} segments. Each segment is maintained as a circular array. So the design of $V$ is similar as $C$ except that its segment is not sorted. Given sample \textit{e}, the segment to keep it is located by $e.id \% m$. The oldest sample in segment $e.id \% m$ will be replaced by \textit{e} when the segment is full. It is  convenient to determine whether an element is in $V$. All the threads in one thread block fetch the samples from a segment and compare them to the element to be checked. Since the number of elements in each segment is the same as the number of threads in the thread block, we only need to do the operation once. Please be noted that not all the visited nodes are pushed into $V$. Otherwise, it would be very expensive to keep all the visited nodes. It is actually unnecessary to do so. The chance of revisiting the early-encountered samples is low as the search moves forward all the way. Similar as~\cite{arxiv22:fabian}, only the nodes used in the expansion are pushed into $V$. This largely avoids the circular search.

The algorithms presented in this section are all implemented in C++ and compiled by GCC \textit{9.3}. The GPU implementation is supported by  NVIDIA CUDA Toolkit \textit{11.2}. We have made it available as supplementary material of this submission.

\section{Experiments}
\label{sec:exp}
\subsection{Experiment Setup}
In this section, the performance of the proposed graph TSDG and NN search approaches is studied in comparison to the state-of-the-art approaches. In particular, our study focuses on graph-based approaches as they are the most effective ones in the literature. For the comparative evaluation with other types of approaches, readers are referred to~\cite{li2019approximate,tbd21:zhao}. Six real-world datasets are adopted to in the NN search evaluation. The brief information about these datasets is summarized in~\autoref{tab:datasets}. As shown on the table, the experiment covers traditional image feature, deep image features, text features, and cross model features. They are all dense and in high dimensions. Different distance measures are adopted for different datasets.  We use local intrinsic dimensionality~\cite{amsaleg2019intrinsic} (LID shown in the \textit{3rd} column) to measure the difficulty of a dataset. Generally, it is more challenging to perform NN search on the datasets with high intrinsic dimensionality.  On the Text-to-Image (T2I) dataset, the local intrinsic dimensionality of the candidate set and the queries is different as they are extracted from different sources. They are \textit{20.9} and \textit{15.5} for text and image features respectively. 

\begin{table}[t]
	\caption{Summary on Datasets used for Evaluation}
	\small{
		\begin{tabular}{lrccl}
			\toprule
			Name                             & $d$   & LID~\cite{amsaleg2019intrinsic} & m($\cdot$,$\cdot$) & Type                            \\
			\midrule
			SIFT~\cite{jegou2010product}     & $128$ & 15.6                            & L2                 & SIFT~\cite{lowe2004distinctive} \\
			DEEP~\cite{babenko2016efficient} & $96$  & 15.9                            & L2                 & Deep                            \\
			GIST~\cite{douze2009evaluation}  & $960$ & 25.9                            & L2                 & GIST~\cite{douze2009evaluation} \\
			GloVe~\cite{pennington2014glove} & $100$ & 29.4                            & Cos                & Text~\cite{pennington2014glove} \\
			SPACEV~\cite{ChenW18}            & $100$ & 23.2                            & L2                 & Text                            \\
			T2I~\cite{t2i2021dataset}        & $200$ & 20.9/15.5                       & IP                 & Text/Image                      \\
	
			\bottomrule
		\end{tabular}
	}
	\label{tab:datasets}
\end{table}

The top-\textit{10} (\textit{Recall@10}) and top-\textit{100} (\textit{Recall@100}) recalls on each dataset are studied under different metrics such as $L2$, \textit{Cosine} and \textit{Inner Product}. Given function $R(i,k)$ returns the number of truth-positive neighbors at top-\textit{k} NN list of sample $i$, the recall at top-$k$ on the whole set is given as
\begin{equation}
	Recall@k=\frac{\sum_{i=1}^n{R(i,k)}}{n{\times}k}.
\label{eval:recall}
\end{equation}

All the experiments are carried out on a machine with one Intel Xeon Processor W-2123 (3.60 Ghz) and 32 GB of memory. The GPU we use is an NVIDIA GeForce RTX 3090. All the codes of different approaches considered in this study are compiled by CUDA Toolkit 11.2 and GCC \textit{9.3}. The optimizations like SIMD and pre-fetching instructions are enabled in the source codes for NN search task. HNSW in Hnswlib~\footnote{https://github.com/nmslib/hnswlib} is treated as the comparison baseline. When we evaluate the performance on the CPU, it runs as a single thread. While for the evaluation on the GPU, it runs in multiple threads.

\begin{table}[t]
	\caption{Time cost (s) spent for Graph Diversification on an existing \textit{k}-NN graph}
	\small{
	\begin{tabular}{llrrrrrr}
		\toprule
		Dataset  & \textit{k} & TSDG  & GD    & NSG~\cite{fu2019fast} & SSG~\cite{fu2021high} & DPG~\cite{li2019approximate} \\
		\midrule
		SIFT1M   & 200 & 54.6  & 18.8  & 188.9                & 47.8                 & 166.9                       \\
		DEEP1M   & 200 & 48.8  & 16.6  & 178.3                & 44.5                 & 133.2                       \\
		GIST1M   & 400 & 234.9 & 167.9 & 1466.2               & 304.2                & 4690.7                      \\
		GloVe1M  & 400 & 142.3 & 49.3  & 1540.8               & 257.2                & 663.3                       \\
		SPACEV1M &200 & 145.2 & 21.9  & 478.5                & 162.8                & 165.9                       \\
		T2I1M    & 400 &  335.9 & 53.6  & 1020.6               & 375.8                & 1015.9                      \\
	
		\bottomrule
	\end{tabular}
	}
	\label{tab:constr_time}
\end{table}

\subsection{The efficiency of Two-stage Diversification}
In our first experiment, we show the efficiency of graph diversification of our approach in comparison to other approaches in the literature. For all the approaches we consider here, the graph diversification is conducted on the same \textit{k}-NN graph for the same dataset. Four approaches are considered in our comparison. The size of \textit{k}-NN list for different datasets is referenced from the experiments in NSG~\cite{fu2019fast} and SSG~\cite{fu2021high}. Four representative approaches such as GD~\cite{tbd21:zhao}, NSG~\cite{fu2019fast}, SSG~\cite{fu2021high} and DPG~\cite{li2019approximate} are considered in the comparison. HNSW is not considered since its diversification is conducted online. GD~\cite{tbd21:zhao} actually adopts the same the diversification scheme as HNSW~\cite{malkov2020efficient} except that it is undertaken on a pre-built \textit{k}-NN graph and expanded to an undirected graph. SSG performs the diversification on the direct neighbors and expanded neighbors. The available implementation of DPG is similar to our second stage diversification. All the  \textit{k}-NN graphs used in the experiments are built by the same GPU-based approach~\cite{wang2021fast}.  So the difference between different approaches lies in the diversification strategies applied on the \textit{k}-NN graph. The time costs for all five approaches are shown in~\autoref{tab:constr_time}, while the time costs on the \textit{k}-NN construction step are considered.

As shown from~\autoref{tab:constr_time}, the time costs across most of the datasets of TSDG are only higher than GD, which is comparable to the first stage diversification of our approach. For all six datasets in million scale, the diversification can be fulfilled in several minutes. In contrast, the diversification of NSG and DPG would take a few hours on the same \textit{k}-NN graph. The operations required by DPG are comparable to the operations in our second stage diversification. However, our approach is much more efficient because there are much fewer edges to process owing to the first stage diversification. 

\begin{figure}[t]
	\begin{center}
	\subfigure
    {\includegraphics[width=0.48\linewidth]{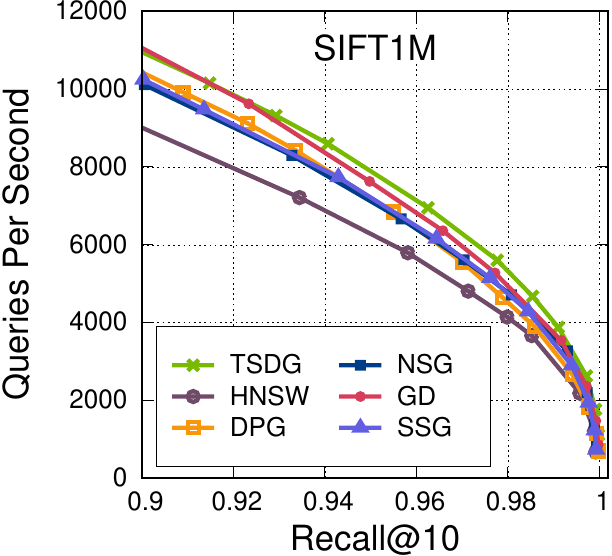}}
    \hspace{0.02in}
	\subfigure
    {\includegraphics[width=0.44\linewidth]{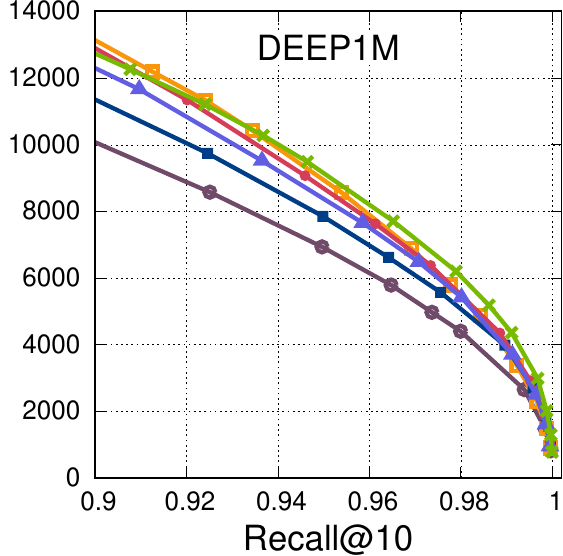}}\\
	\hspace{0.03in}
	\subfigure
	{\includegraphics[width=0.47\linewidth]{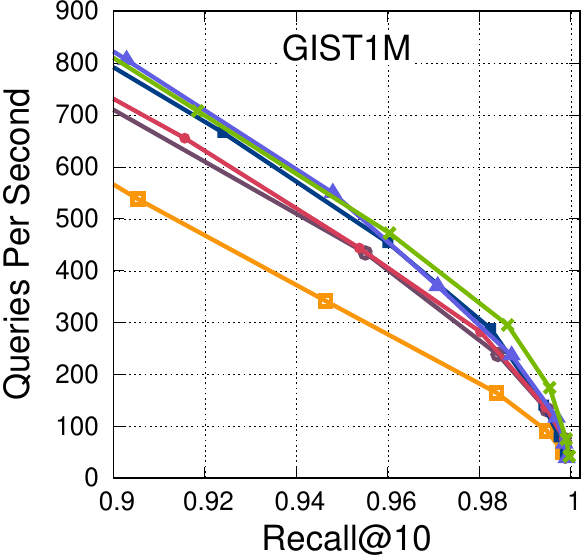}}
    \hspace{0.08in}
	\subfigure
	{\includegraphics[width=0.43\linewidth]{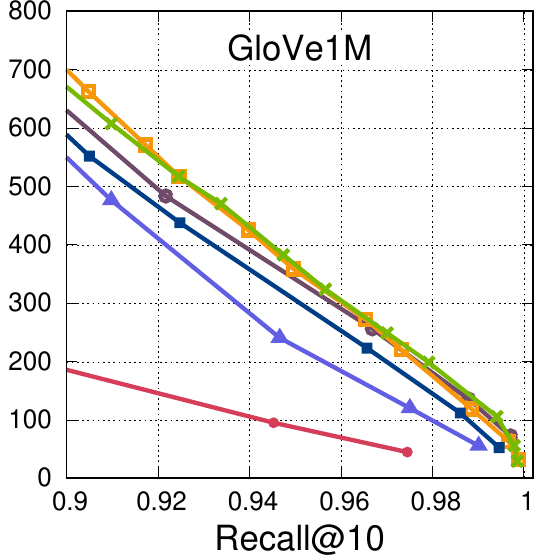}}\\
	\subfigure
	{\includegraphics[width=0.48\linewidth]{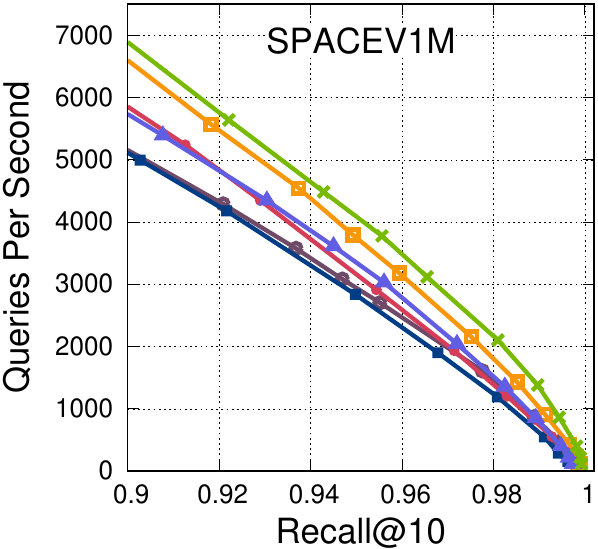}}
    \hspace{0.02in}
	\subfigure
	{\includegraphics[width=0.44\linewidth]{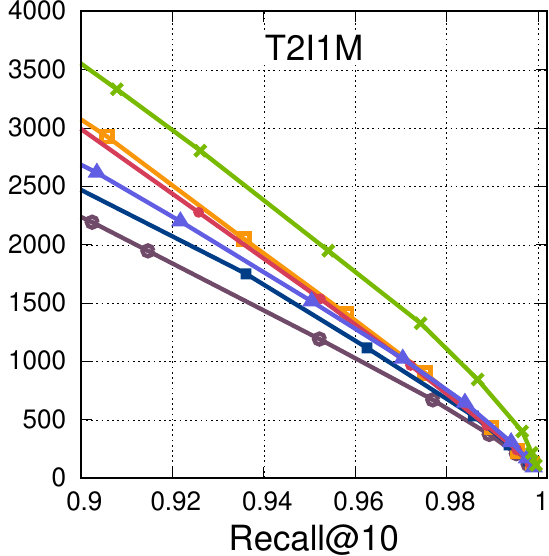}}\\

	\caption{The NN search performance on the CPU. The parameters used in HNSW, NSG and SSG graph index construction follow with the SSG~\cite{fu2021high}.}
	\label{fig:cpu_search}
\end{center}
\end{figure}

\begin{figure}[t]
	\begin{center}
		\subfigure[SIFT1M-BS1]
		{\includegraphics[width=0.47\linewidth]{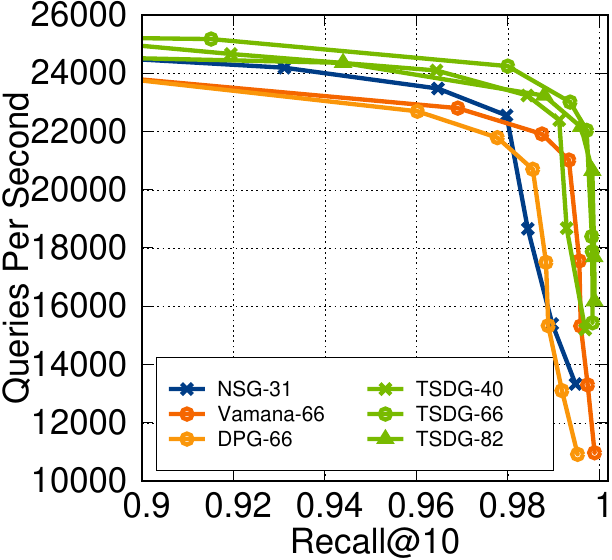}}
		\hspace{0.02in}
		\subfigure[SIFT1M-BS10]
		{\includegraphics[width=0.465\linewidth]{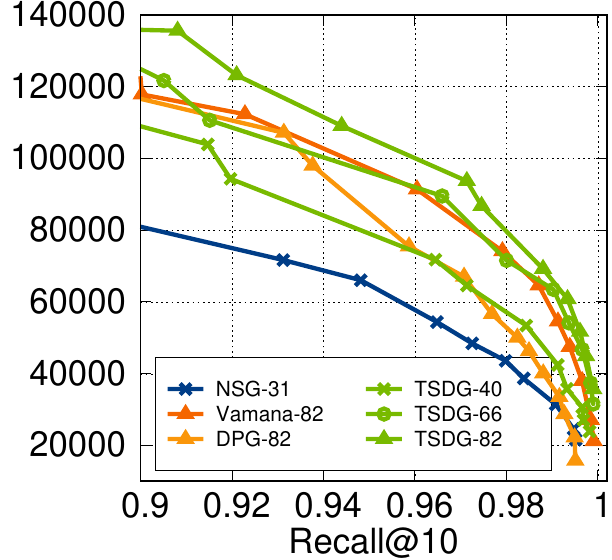}}\\
		\subfigure[SIFT1M-BS100]
		{\includegraphics[width=0.49\linewidth]{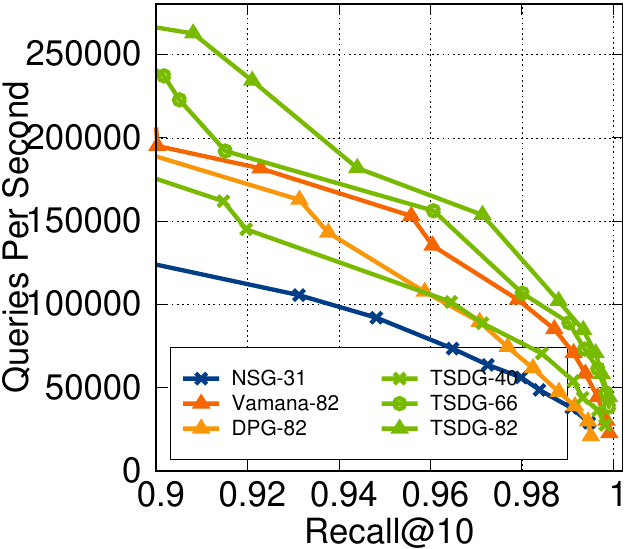}}
		\subfigure[SIFT1M-BS10k]
		{\includegraphics[width=0.48\linewidth]{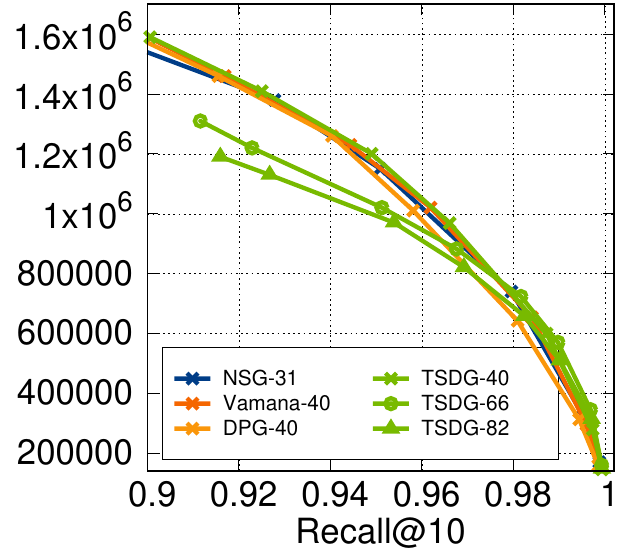}}\\
	
		\caption{The performance of NN search with the support of graphs in different average degree and batch size. The numbers in the label indicate the average degree of one graph. ``SIFT1M-BSx'' is the run queries arrive in batch size of \textit{x}.}
		\label{fig:gpu_degree}
	\end{center}
\end{figure}	

\subsection{NN Search Performance on the CPU}
With the support of built graph, we proceed to study the NN search performance of different approaches. In this experiment, we adopt the same NN search implementation from NSG code~\footnote{https://github.com/ZJULearning/nsg/} with additional \textit{32} random starting seeds for TSDG, GD, NSG, SSG, and DPG. So the major difference among them lies in the graph that supports the NN search. HNSW is adopted as the comparison baseline. We consider the curve of Recall@10 against the number of queries that one approach processes in one second.
For clarity, our NN search approach that is supported by the proposed TSDG graph is given as ``TSDG-BSx''. ``BSx'' indicates the batch size of a search task. For instance, ``BS10'' indicates the batch size of the search task is \textit{10}.  The search performance on the six million-scale datasets is shown in~\autoref{fig:cpu_search}.

As shown from the figures, NN search supported by TSDG outperforms the rest on all six datasets. A considerable performance gap is observed on challenging datasets such as T2I1M and GloVe1M between TSDG and the rest. DPG performs well on \textit{4} out of \textit{6} datasets. It shows inferior performance to TSDG when the data dimension is high. TSDG shares similar average graph degree as that of DPG. Compared to DPG, our graphs keep more informative edges owing to the first stage diversification. As a result, the NN search reaches nearest neighbors with fewer comparisons.


\subsection{NN Search Performance on the GPU}
In this section, we continue to study the performance of our NN search approaches supported by our graph on the GPU. Before we proceed, we show the impact of the connectivity of a graph on the search performance particularly for small batch queries. The connectivity of a graph is measured by its average degree. In the following, we study the performance trend for small batch queries and large batches on the GPU. Approach NSG with average graph degree of \textit{31} is considered as the comparison baseline. Vamana~\cite{subramanya2019diskann} is a diversifying approach only adopts the first stage graph diversification in ours. While DPG~\cite{li2019approximate} is comparable to the diversification of our second stage. Different runs of our approach are carried out supported by graphs with different average degrees. The same NN search that is presented in~\autoref{alg:simple_search} is used on all the graphs for batch sizes \textit{1}, \textit{10} and \textit{100}. While~\autoref{alg:common_search} is used on all the graphs for batch size \textit{10k}. The experiments are conducted on SIFT1M only for page limit. Similar trend is observed on other datasets.
	
\begin{figure}[t]
	\begin{center}
		\subfigure
		{\includegraphics[width=0.49\linewidth]{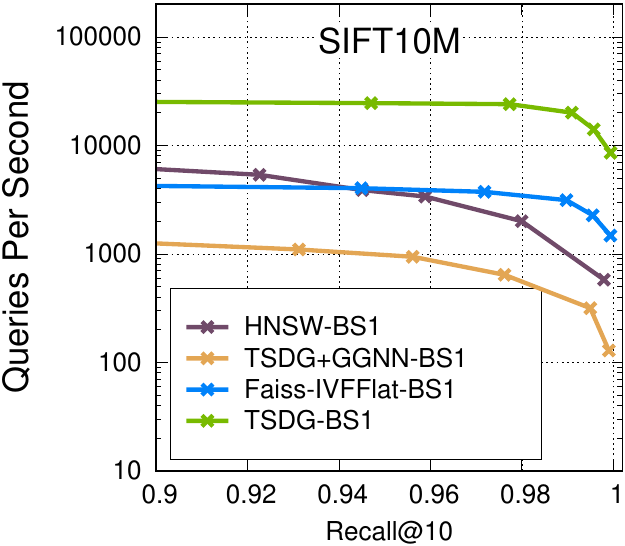}}
		\subfigure
		{\includegraphics[width=0.455\linewidth]{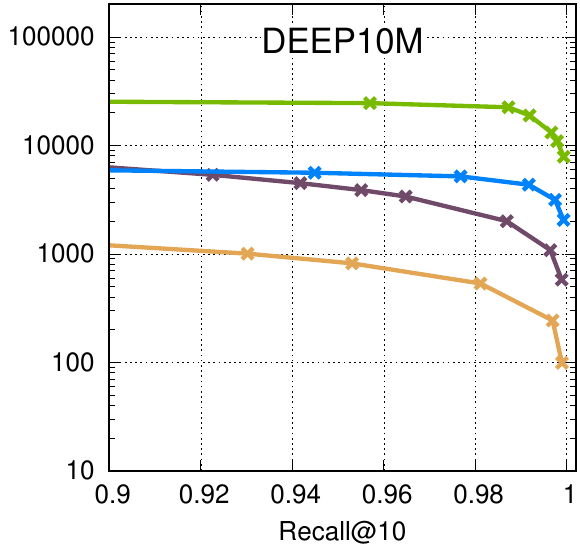}} \\
		\subfigure
		{\includegraphics[width=0.49\linewidth]{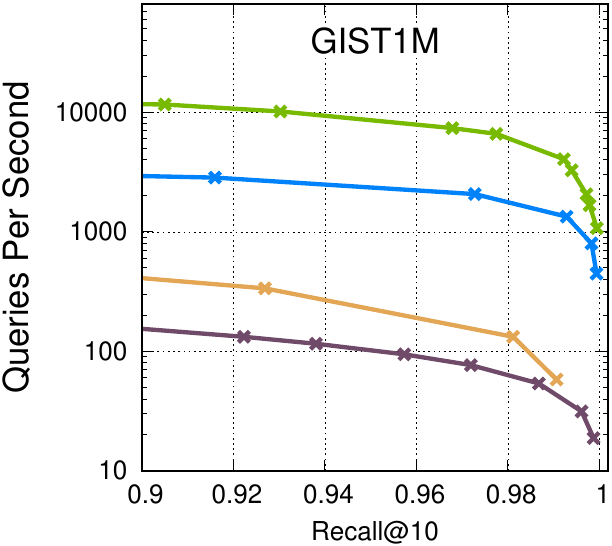}}
		\subfigure
		{\includegraphics[width=0.46\linewidth]{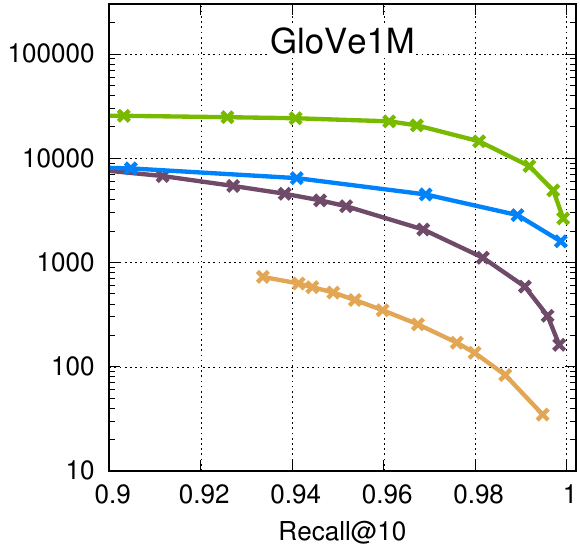}} \\
		\subfigure
		{\includegraphics[width=0.49\linewidth]{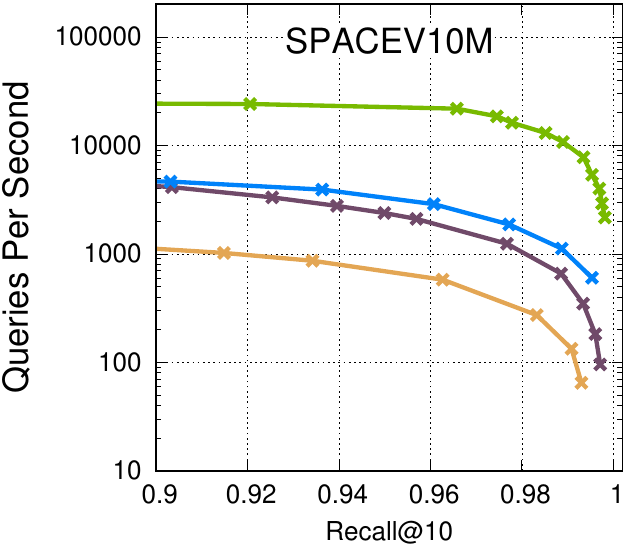}}
		\subfigure
		{\includegraphics[width=0.445\linewidth]{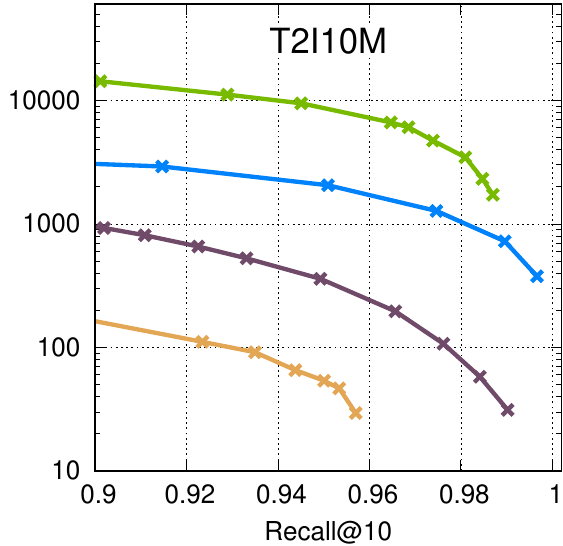}}
		\caption{The NN (\textit{k}=\textit{10}) search performance on the GPU in batch size \textit{1}.}
		\label{fig:gpu_search_bs1_k10}
	\end{center}
\end{figure}

\begin{figure}[t]
	\begin{center}
		\subfigure
		{\includegraphics[width=0.49\linewidth]{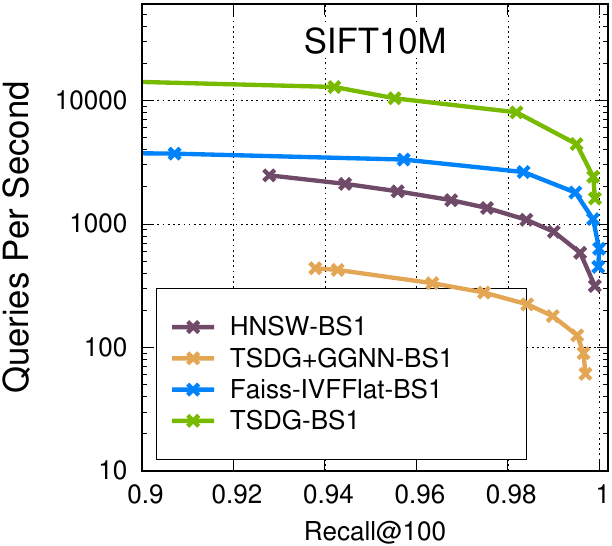}}
		\subfigure
		{\includegraphics[width=0.455\linewidth]{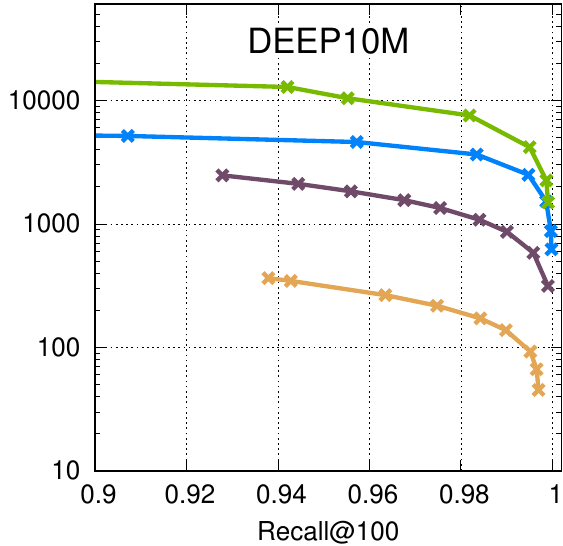}} \\
		\subfigure
		{\includegraphics[width=0.49\linewidth]{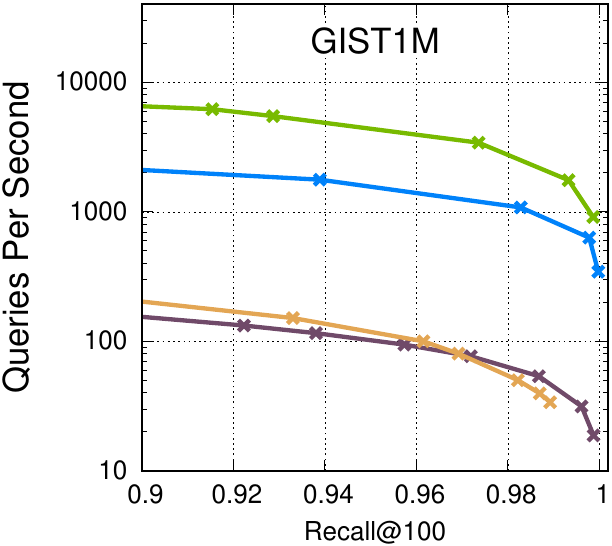}}
		\subfigure
		{\includegraphics[width=0.46\linewidth]{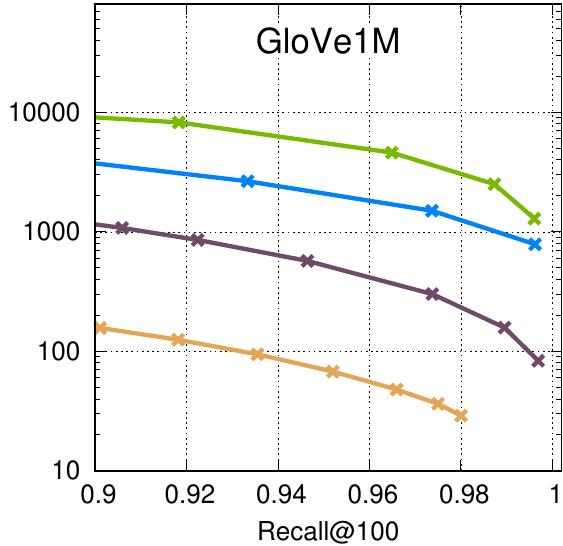}} \\
		\subfigure
		{\includegraphics[width=0.49\linewidth]{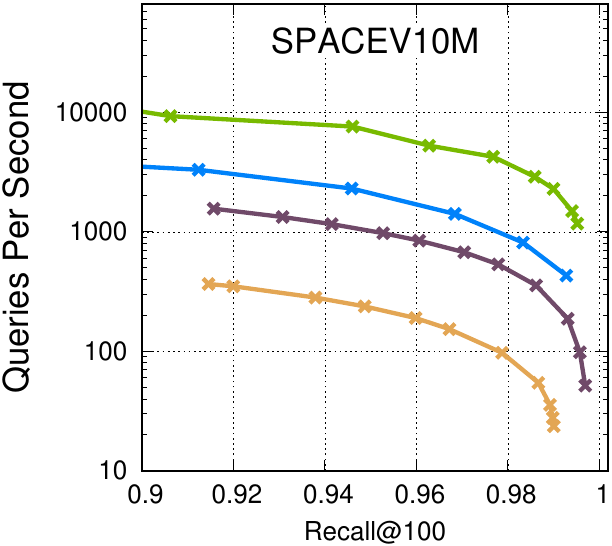}}
		\subfigure
		{\includegraphics[width=0.445\linewidth]{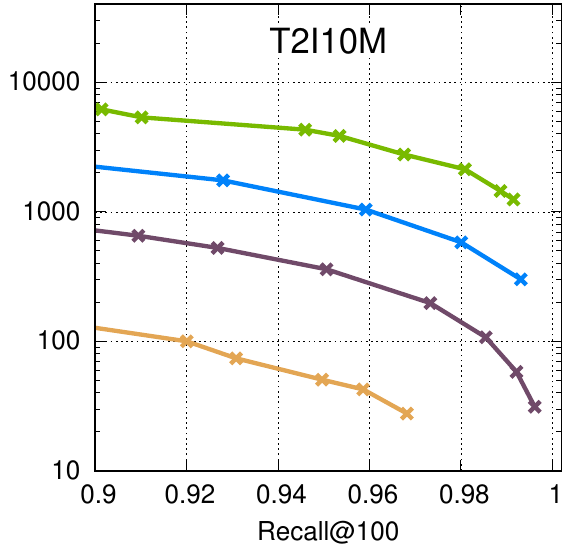}}
		\caption{The NN (\textit{k}=\textit{100}) search performance on the GPU in batch size \textit{1}.}
		\label{fig:gpu_search_bs1_k100}
	\end{center}
\end{figure}

\begin{figure*}[t]
	\begin{center}
		\hspace{0.3in}
		\includegraphics[width=0.65\linewidth]{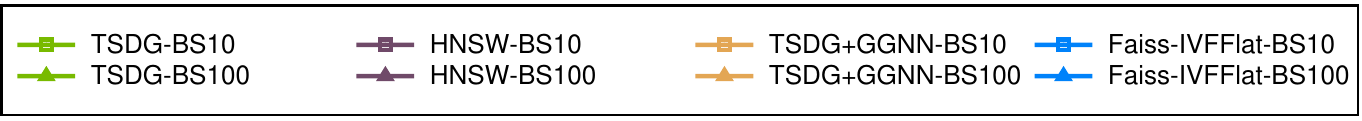}
	\end{center}
\end{figure*}

\begin{figure}[t]
	\begin{center}
		\vspace{-0.15in}
		\subfigure
		{\includegraphics[width=0.49\linewidth]{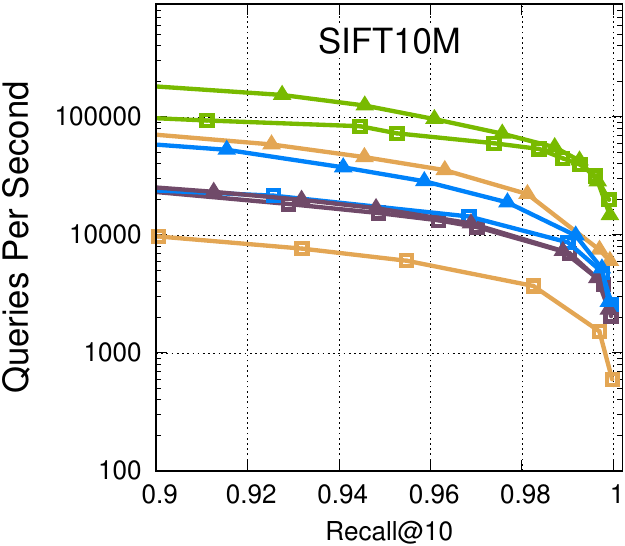}}
		\subfigure
		{\includegraphics[width=0.455\linewidth]{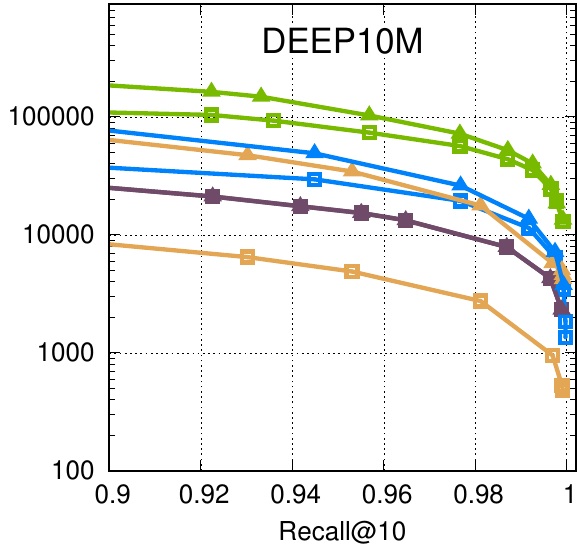}} \\
		\subfigure
		{\includegraphics[width=0.48\linewidth]{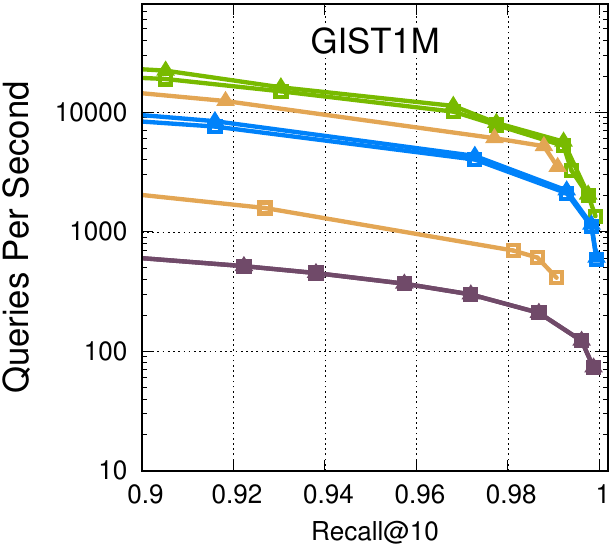}}
		\subfigure
		{\includegraphics[width=0.45\linewidth]{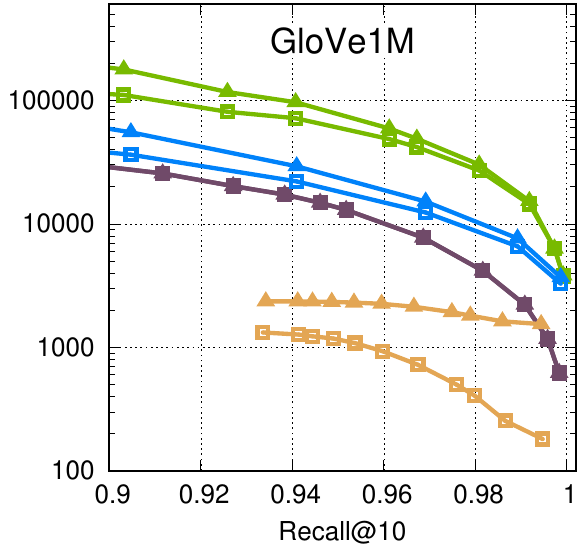}} \\
		\subfigure
		{\includegraphics[width=0.49\linewidth]{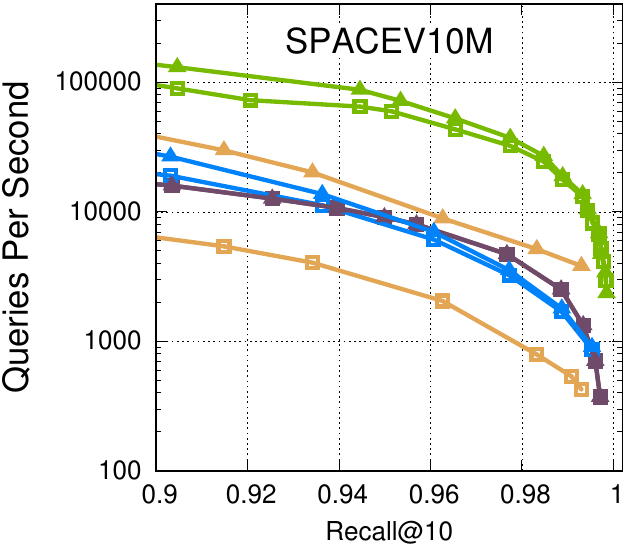}}
		\subfigure
		{\includegraphics[width=0.445\linewidth]{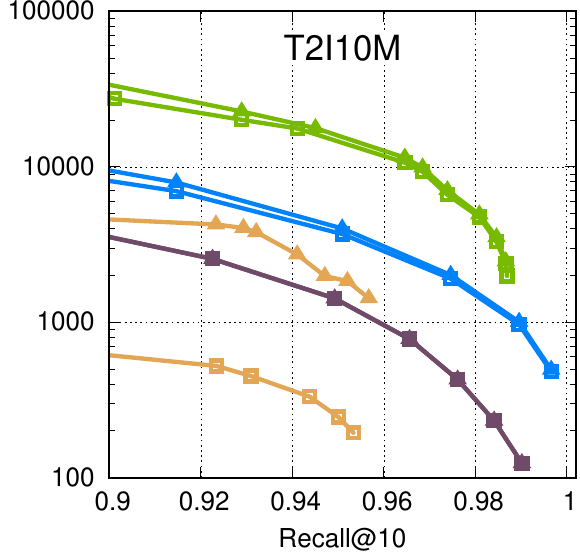}}
		\caption{The NN (\textit{k}=\textit{10}) search performance on the GPU in batch size \textit{10} and \textit{100}.}
		\label{fig:gpu_search_bs_small_k10}
	\end{center}
\end{figure}

\begin{figure}[t]
	\begin{center}
		\vspace{-0.15in}
		\subfigure
		{\includegraphics[width=0.49\linewidth]{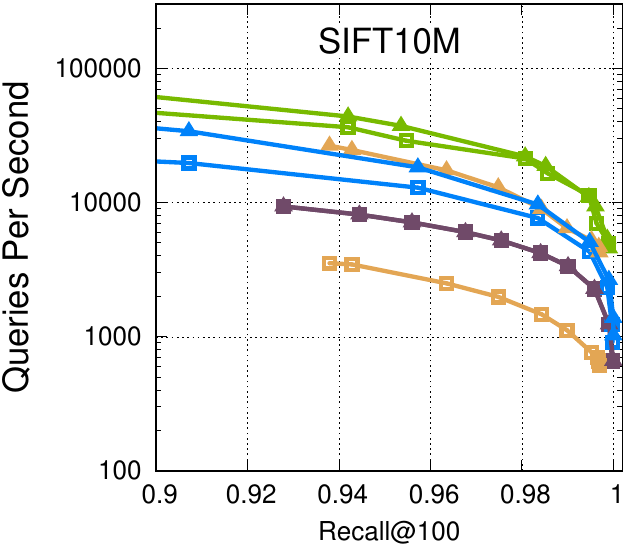}}
		\subfigure
		{\includegraphics[width=0.455\linewidth]{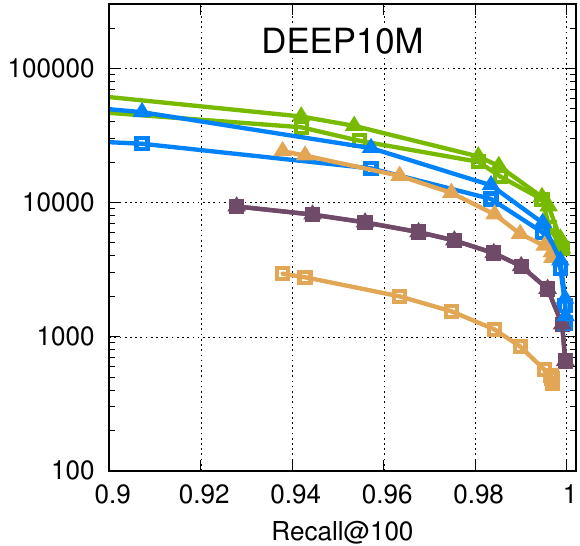}} \\
		\subfigure
		{\includegraphics[width=0.48\linewidth]{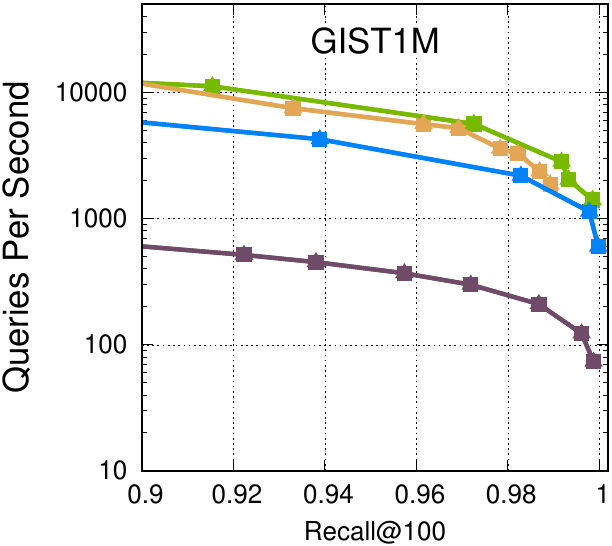}}
		\subfigure
		{\includegraphics[width=0.45\linewidth]{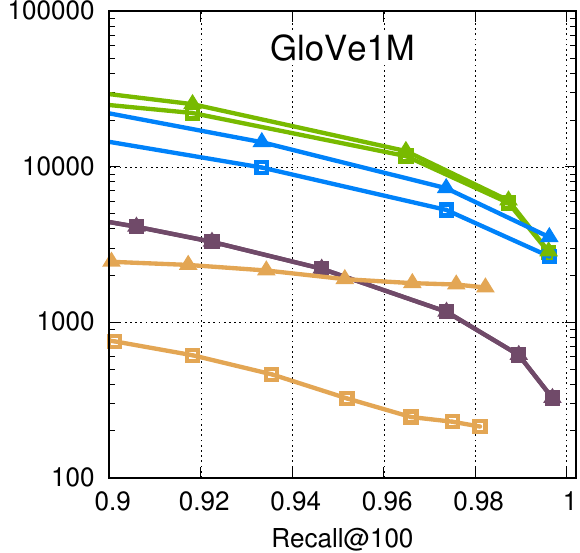}} \\
		\subfigure
		{\includegraphics[width=0.49\linewidth]{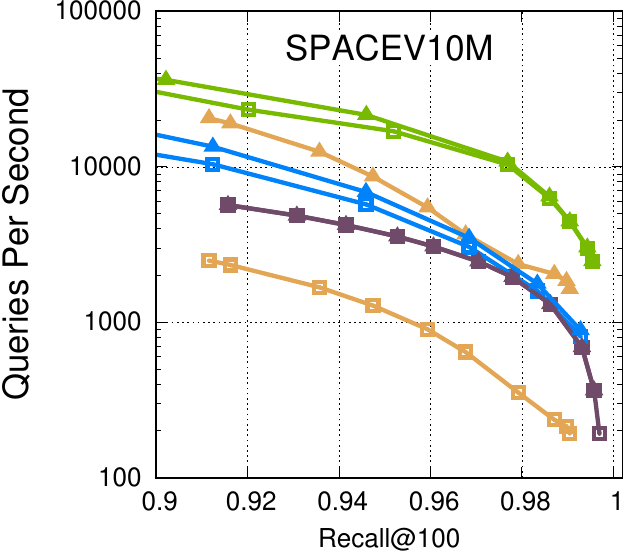}}
		\subfigure
		{\includegraphics[width=0.445\linewidth]{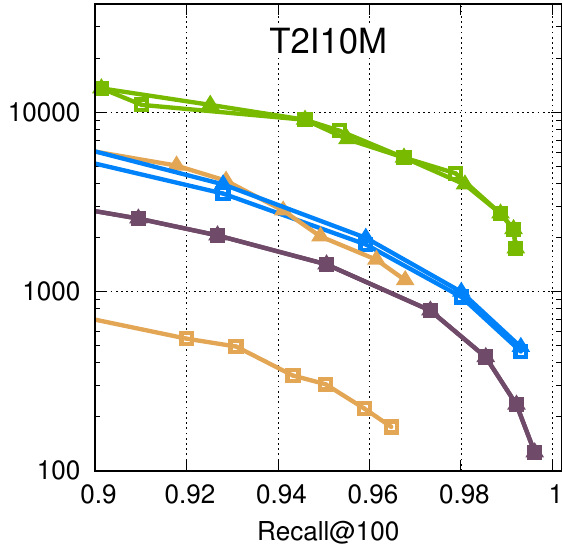}}
		\caption{The NN (\textit{k}=\textit{100}) search performance on the GPU in batch size \textit{10} and \textit{100}.}
		\label{fig:gpu_search_bs_small_k100}
	\end{center}
\end{figure}

\begin{figure}[t]
	\begin{center}
		\subfigure
		{\includegraphics[width=0.49\linewidth]{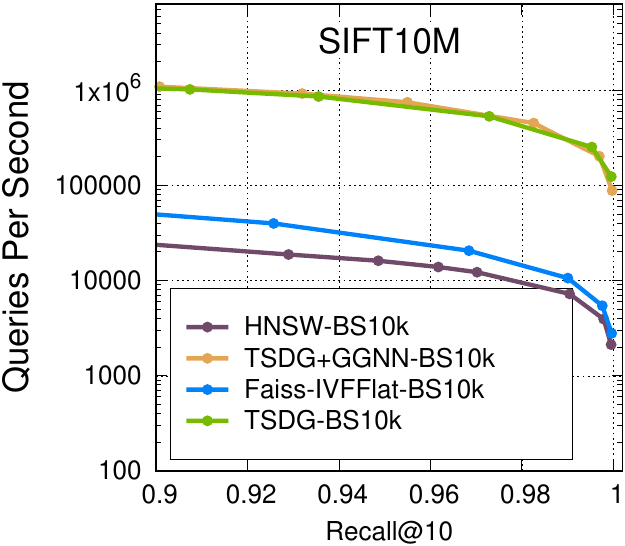}}
		\subfigure
		{\includegraphics[width=0.455\linewidth]{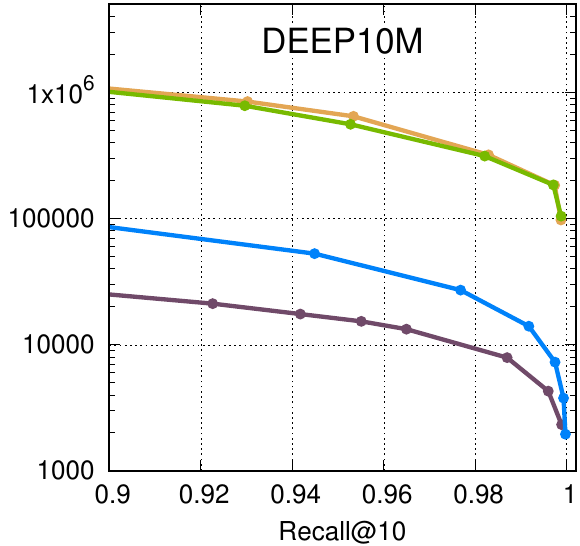}}
		\\
		\subfigure
		{\includegraphics[width=0.49\linewidth]{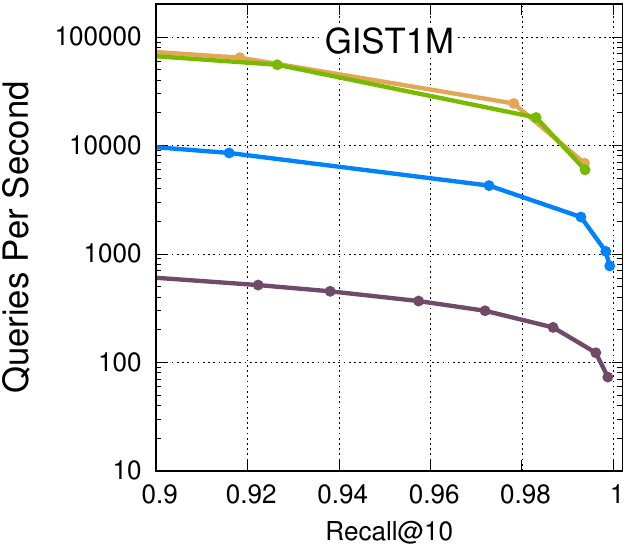}}
		\subfigure
		{\includegraphics[width=0.455\linewidth]{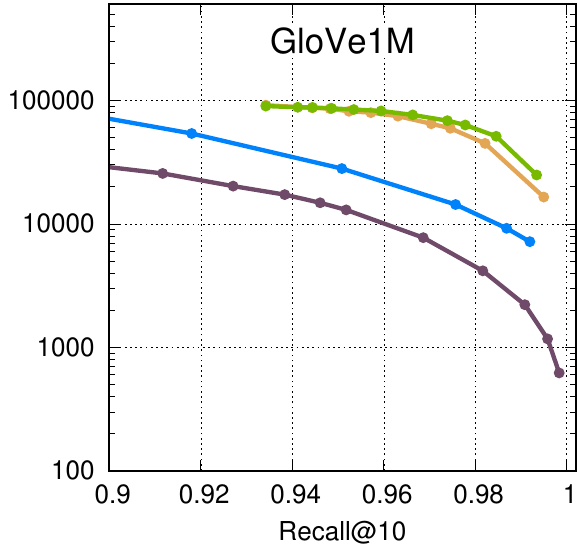}}
		\\
		\subfigure
		{\includegraphics[width=0.49\linewidth]{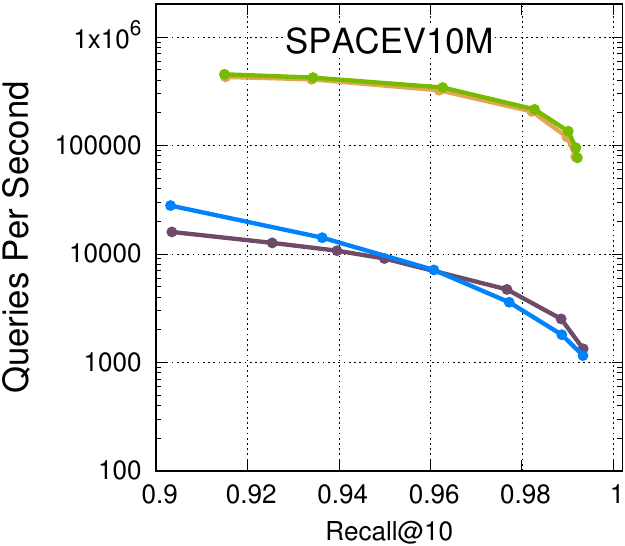}}
		\subfigure
		{\includegraphics[width=0.455\linewidth]{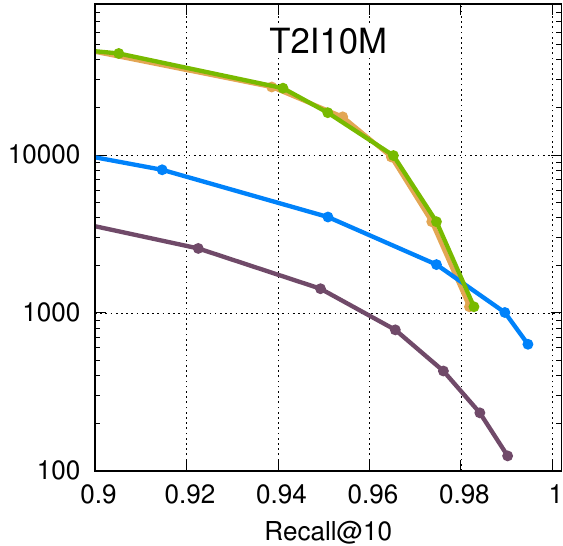}}

		\caption{The NN (\textit{k}=\textit{10}) search performance on the GPU in batch size \textit{10k}.}
		\label{fig:gpu_search_bs10k_k10}
	\end{center}
\end{figure}

\begin{figure}[t]
	\begin{center}
		\subfigure
		{\includegraphics[width=0.49\linewidth]{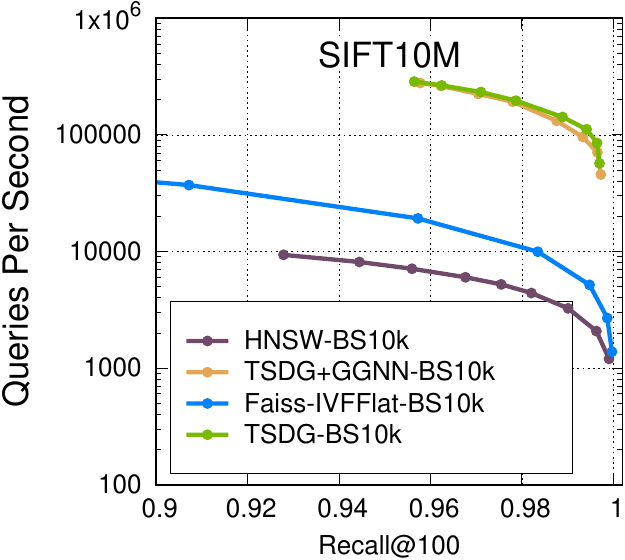}}
		\subfigure
		{\includegraphics[width=0.455\linewidth]{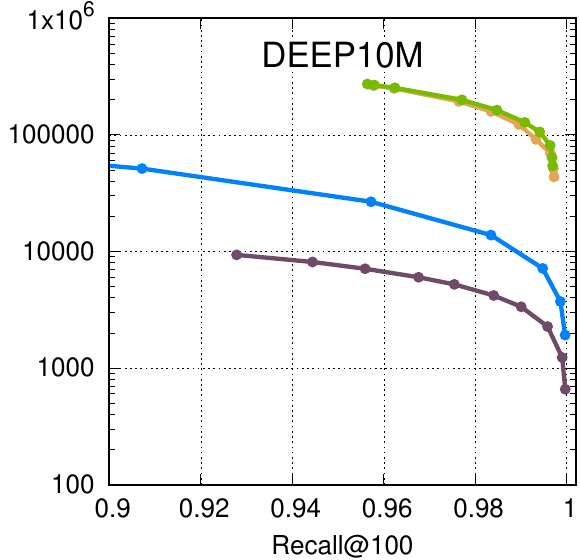}}
		\\
		\subfigure
		{\includegraphics[width=0.49\linewidth]{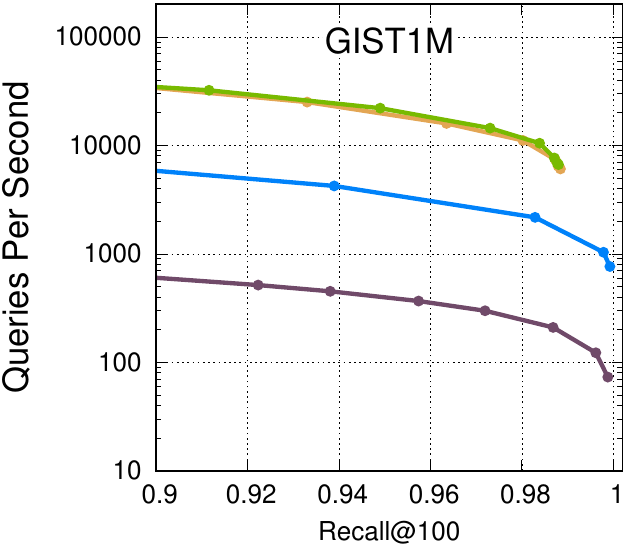}}
		\subfigure
		{\includegraphics[width=0.455\linewidth]{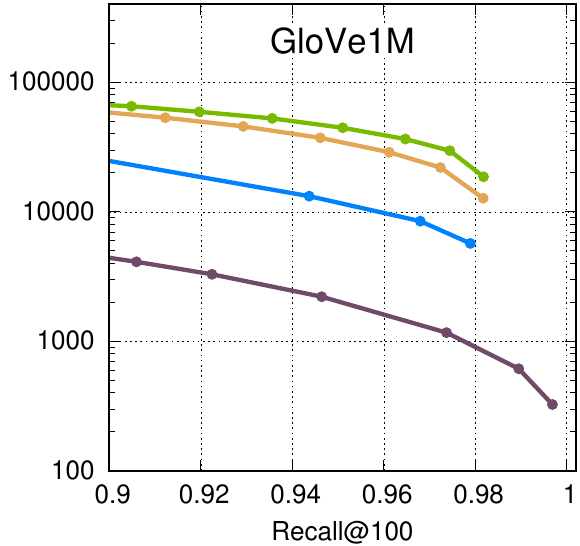}}
		\\
		\subfigure
		{\includegraphics[width=0.49\linewidth]{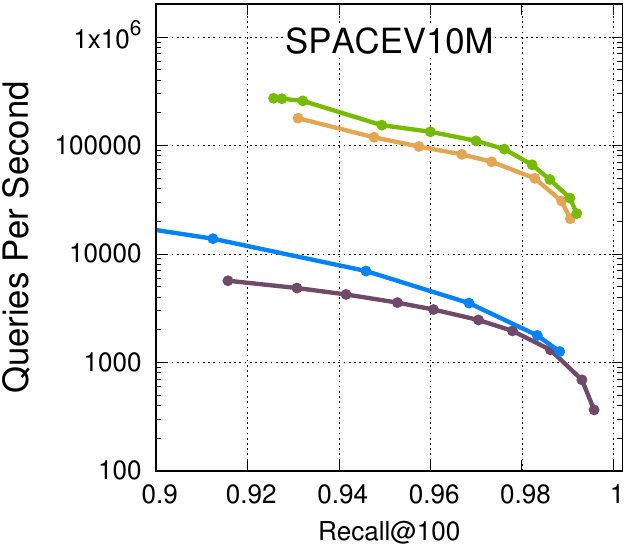}}
		\subfigure
		{\includegraphics[width=0.455\linewidth]{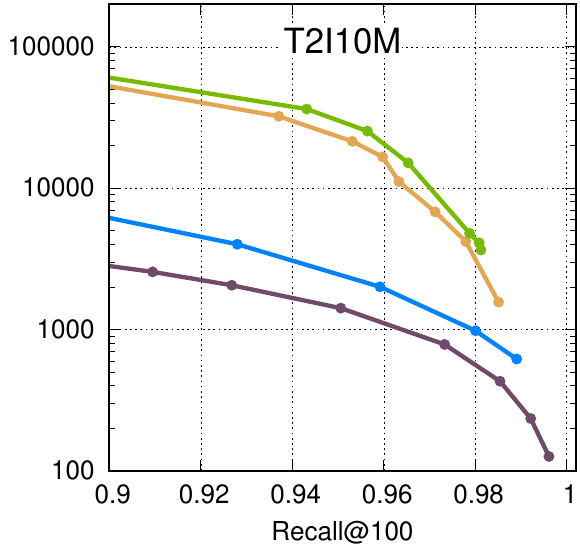}}

		\caption{The NN (\textit{k}=\textit{100}) search performance on the GPU in batch size \textit{10k}.}
		\label{fig:gpu_search_bs10k_k100}
	\end{center}
\end{figure}

As shown from~\autoref{fig:gpu_degree}, higher graph degree leads to considerably better performance for different small batch sizes. The performance trend of large batch searches on each graph is similar to that of on the CPU.  For the graphs with the same average degree, TSDG shows better performance than that of Vamana and DPG. This does indicate the two-stage graph diversification is much more effective than one-stage schemes. In this experiment, our graphs are pruned such that their average degree is in line with Vamana graphs. Such setting is not the optimal for NN search. In the following experiment, we visit the edges whose occlusion factors are lower than \textit{10} for small batch queries. While for large batch queries, we only visit the edges whose occlusion factors are lower than \textit{5}.


The experiment about NN search on the GPU is conducted on six datasets as before. The NN search is scaled up to \textit{10} million level for SIFT, DEEP, SPACEV, and T2I datasets as it is much efficient to build the graphs on the GPU. The queries are organized in both small batches (batch size=\textit{10}) and large batches (batch size=\textit{10k}). In the experiment, the performance from HNSW, GGNN, SONG, and Faiss-IVFFlat is reported. For HNSW, it runs on multiple CPU threads, which is the only approach that runs on the CPU in our study. For approach SONG, we can only run its code\footnote{https://github.com/sunbelbd/song} on SIFT1M smoothly.  For this reason, a separate comparison is made with SONG on SIFT1M alone. We re-implement the algorithm of GGNN with our framework and graphs, which performs much better than the original authors' implementation. This run is given as ``TSDG-GGNN''. The NN search results are shown as the curve of recall against the number of processed queries in one second. Results from the same approach are shown with curves in the same color. Results in the same query batch size are shown with curves with the same decoration. 

According to our observation on SIFT1M dataset, the performance of our NN search is significantly better than SONG for all batch sizes. When the batch size is \textit{1}, our approach is faster than SONG by more than \textit{20} times, which only could process less than \textit{1000} queries per second under the same search recall. As the batch size increases, the performance gap between our approach and SONG decreases. However, it is still around two times faster than SONG when batch size is \textit{10k}. This observation is in line with~\cite{arxiv22:fabian}. As the implementation of SONG is difficult to scale-up to larger datasets, it is not involved in following experiments.

\autoref{fig:gpu_search_bs1_k10} and~\autoref{fig:gpu_search_bs1_k100} show NN search efficiency of \textit{Recall@10} and \textit{Recall@100} respectively for our approach in comparison to HNSW, TSDG+GGNN, and Faiss-IVFFlat when the batch size is only one. Our NN search demonstrates several times higher performance than the rest when only one query arrives in each batch. This owes mainly to the specific design of the search procedure for the small batch case, which utilizes the GPU parallelism apparently better. Although the performance gap between ours and the rest on \textit{Recall@100} goes narrower, the advantage of our approach is still significant. Similar observation holds when the batch size increases from \textit{1} to \textit{100}. As shown in~\autoref{fig:gpu_search_bs_small_k10} and~\autoref{fig:gpu_search_bs_small_k100}), our NN search still performs much better than the rest. Most of the approaches except for GGNN shows no considerable performance boost when the batch size increases, although better utilization of parallelism is expected for all of them. Although the performance of HNSW could benefit from more CPU cores, the performance of HNSW is quickly saturated as the batch size increases because the bandwidth of the CPU memory is far less than that of the GPU memory.

The NN search results on \textit{Recall@10} and \textit{Recall@100} of our approach on large batch size (\textit{10k}) are shown in \autoref{fig:gpu_search_bs10k_k10} and \autoref{fig:gpu_search_bs10k_k100} respectively, in comparison to HNSW, TSDG+GGNN, and Faiss-IVFFlat. In this case, the performance of our approach is similar to GGNN when we examine top-10 recall (\autoref{fig:gpu_search_bs10k_k10}).  Nevertheless, when we check \textit{Recall@100}, we find our approach outperforms GGNN considerably on challenging datasets such as SPACEV10M and T2I10M. As shown in~\autoref{fig:gpu_search_bs10k_k100}, it is consistently better than GGNN and shows \textit{40\%} improvement over GGNN to its best on challenging datasets GloVe1M, SPACEV10M and T2I10M. This owes to the data structures we designed that significantly reduce the maintenance cost when they are in large size. According to our offline test, GGNN shows more than \textit{10\%} lower performance if it is supported the graph proposed in the original paper. From this sense, it is clear to see the proposed TSDG is also supportive for other graph-based search procedures.

As seen from all the above experiments, the NN search by our approach shows the best performance across all the datasets under different circumstances. Nevertheless, similar to all other approaches, the performance drops steadily as intrinsic dimensionality of the dataset increases. If we compare the performance on GloVe1M and SPACEV1M (\autoref{fig:cpu_search}(d) and (e)), we find that it is much more efficient to perform NN search on SPACEV1M than that of GloVe1M although they are on the same scale and the same dimensionality. This is where the intrinsic dimensionality comes to play. Dataset GloVe1M is more challenging since the data are distributed on a manifold with the higher intrinsic dimensionality. In such case, there are more samples that share similar distances to the query at each iteration. The NN search has to compare to more potential samples on the search path before it reaches the nearest neighbor.

\section{Conclusion}
In this paper, we have optimized the whole process of the graph-based NN search, namely the NN search procedure and the graph that supports the NN search. The graph is built by a two-stage diversification on the \textit{k}-NN graph. Compared with the state-of-the-art approaches, our graph is not only efficient to be built, but also leads to the best search performance across different datasets on the CPU. Moreover, the graph is flexible to fit with different search scenarios where the search tasks and the available computing resources vary. Accordingly, two highly efficient NN search algorithms have been designed to make the full use of GPU computing power. Inside each of the search algorithm, two key data structures have been carefully designed. The design well addresses the bandwidth mismatch between the GPU cores and the GPU memories. When our search algorithms are supported by the proposed graph, they  outperform considerably the other NN search approaches on the GPU, particularly for queries in small batches.


\begin{acks}
  This work is supported by National Natural Science Foundation of China under grants 61572408 and 61972326, and the grants of Xiamen University 20720180074.
\end{acks}

\bibliographystyle{ACM-Reference-Format}
\bibliography{sigir22}

\end{document}